\documentclass[manuscript,screen]{acmart}
\usepackage{algorithmic}
\usepackage{algorithm}
\usepackage{array}
\usepackage{stfloats}
\usepackage{graphicx}
\usepackage{enumitem}
\usepackage{xcolor}
\usepackage{booktabs}
\usepackage{pifont}
\usepackage{multirow}

\newcommand{\rqbox}[1]{%
  \par\smallskip
  \noindent
  \setlength{\fboxsep}{5pt}
  \colorbox{gray!25}{%
    \begin{minipage}{\dimexpr\columnwidth-2\fboxsep\relax}
    \setlength{\parindent}{0pt}%
    \setlength{\parskip}{0.3em}%
    #1
    \end{minipage}%
  }%
  \par\smallskip
}
\AtBeginDocument{%
  }

\setcopyright{cc}
\setcctype{by}
\copyrightyear{2026}
\acmDOI{XXXXXXX.XXXXXXX}
\acmJournal{TOSEM}
\acmYear{2026} \acmVolume{1} \acmNumber{1} \acmArticle{}
\acmMonth{1}
\begin{document}

\title{Faster Code, Deeper Debt? A Multivocal Literature Review on Technical Debt \textcolor{black}{and Its Early Signs} in LLM-Assisted Software Development}

\renewcommand{\shorttitle}{Faster Code, Deeper Debt? A Multivocal Literature Review}

\author{Ramtin Ehsani}
\email{ramtin.ehsani@drexel.edu}
\affiliation{%
  \institution{Drexel University}
  \city{Philadelphia}
  \state{Pennsylvania}
  \country{USA}
}

\author{Shriya Rawal}
\email{sr3728@drexel.edu}
\affiliation{%
  \institution{Drexel University}
  \city{Philadelphia}
  \state{Pennsylvania}
  \country{USA}
}

\author{Yuanfang Cai}
\email{yc349@drexel.edu}
\affiliation{%
  \institution{Drexel University}
  \city{Philadelphia}
  \state{Pennsylvania}
  \country{USA}
}

\author{Preetha Chatterjee}
\email{preetha.chatterjee@drexel.edu}
\affiliation{%
  \institution{Drexel University}
  \city{Philadelphia}
  \state{Pennsylvania}
  \country{USA}
}





\renewcommand{\shortauthors}{Ehsani et al.}

\begin{abstract}
With the rapid adoption of LLM-assisted coding, the need to manage the technical debt these systems introduce has become urgent. In this paper, we conduct a multivocal literature review of \textcolor{black}{104 sources (31 formal, 73 grey)} to examine how LLM-assisted development contributes to technical debt and what strategies, metrics, and benchmarks exist to mitigate it.
We find that LLMs often amplify traditional forms of technical debt, particularly code, design, and documentation debts, while also introducing new LLM-specific debts. Notably, we identify fast-integration debt, where rapidly generated code prioritizes speed over quality, triggering a domino effect that leads to governance debt and increased long-term maintenance costs. Additional emerging categories include prompt, ethical, data, and provenance debt, reflecting new challenges unique to LLM adoption. 
To address these, strategies suggested in the literature include human-in-the-loop frameworks, prompt engineering, and data quality alignment. In practice, tools such as SonarQube are commonly used to detect technical debt indicators, while research prototypes such as CodeSmellEval are emerging to assess how LLMs contribute to debts. However, no standardized benchmarks or LLM-specific metrics yet exist, leaving an important gap. Based on findings, we outline insights and future directions to ensure reliable integration of LLMs into software engineering workflows.
\end{abstract}



\received{20 February 2007}
\received[revised]{12 March 2009}
\received[accepted]{5 June 2009}

\maketitle

\section{Introduction}
Large language models (LLMs) are rapidly transforming how developers write and maintain code. These tools promise increased productivity and reduced development effort~\cite{10.1145/3756681.3757081}, yet their long-term implications, particularly regarding technical debt, remain insufficiently understood. Developers are increasingly relying on LLMs for quick fixes, boilerplate, and even entire modules, but the resulting code may suffer from inconsistency, poor testing, limited documentation, or reduced maintainability~\cite{velasco_how_2025, ribeiro2026llmsinternalrepresentationcode, valentin2025incoherenceoraclelessmeasureerror}. These potential risks warrant closer attention as LLMs become a regular part of software development practice, especially with the emerging vision of AI-native Software Engineering (SE 3.0), where human-AI collaborations become more integrated into the development workflows~\cite{li2025riseaiteammatessoftware, hassan2024ainativesoftwareengineeringse}.

\textcolor{black}{
Emerging evidence points to a possible trade-off between short-term productivity gains and long-term software maintenance efforts. 
While several studies suggest that AI tools can boost developer productivity~\cite{github_study_2022,11121735, peng2023impactaideveloperproductivity}, their broader implications for long-term maintenance remain unclear. For example, a large-scale analysis by GitClear, which analyzed 211 million lines of code from major technology companies between 2020 and 2024, reported increased code duplication, short-term churn, and a sharp decline in code reuse \cite{gitclear_ai_2025}. For the first time, ``copy/paste'' style duplication outpaced refactored code reuse, signaling a shift toward practices that may accelerate the accumulation of technical debt \cite{gitclear_ai_2025}. Because these issues often do not cause immediate failures, they can be easily overlooked during developer reviews~\cite{liu2026debtaiboomlargescale}.} 

\begin{figure*}[h!]
    \centering
    \includegraphics[width=0.65\textwidth]{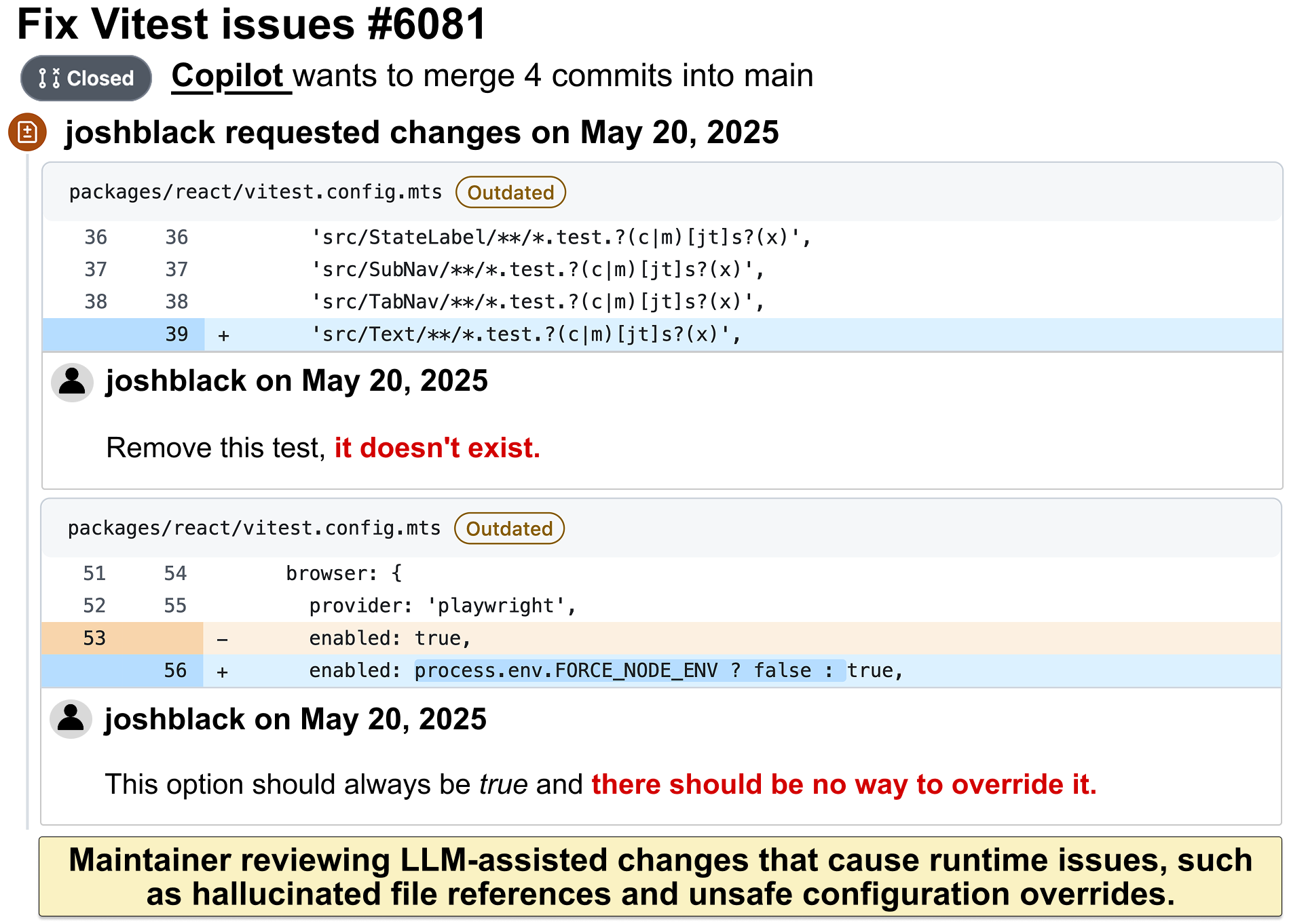}
    \caption{\textcolor{black}{Example of LLM-assisted code (by Copilot) submitted to project \textit{primer/react} on GitHub (3.8k stars), highlighting new risks such as hallucinated file references and unsafe configuration overrides.}}
    \label{fig:intro_example}
\end{figure*}

\textcolor{black}{LLM-generated code introduces risks that differ in important ways from those of traditional human-written code. For instance, Figure~\ref{fig:intro_example} shows a pull request involving LLM-assisted code generation using GitHub Copilot\footnote{\url{https://github.com/primer/react/pull/6081}}. The generated code includes hallucinated file references and unsafe configuration overrides, highlighting new risks that are unique to LLM-assisted development workflows~\cite{liu2026debtaiboomlargescale}. 
LLM-generated code can also be highly prompt-dependent. 
Small changes in prompts or model settings (e.g., higher temperature) can produce different implementations for the same functionality, making system behavior harder to reproduce~\cite{vangala2026aigeneratedcodereproducibleyet}. 
Moreover, prompts and inputs to LLMs themselves become first-class development artifacts, introducing new challenges in how they are preserved, versioned, and maintained over time~\cite{11025760}.
Unlike human-written code, which reflects deliberate design decisions and traceable authorship, LLM-generated code is often produced non-deterministically. This can lead developers to integrate code whose logic or assumptions are not fully understood, increasing the cognitive effort required to evaluate, maintain, and eventually repay the technical debts it may introduce~\cite{storey2026technicaldebtcognitiveintent}. 
These potential risks and their early signals must be understood more thoroughly to support the reliable integration of LLMs into software engineering workflows.} 

\textcolor{black}{While prior work has examined correctness and quality issues in LLM-generated code~\cite{10.5555/3666122.3667065, 11025576}, these perspectives primarily capture local or task-specific software quality concerns, but do not fully explore how such issues could accumulate into long-term technical debt. 
This distinction is critical in the context of LLM-assisted development, where the literature is still emerging and explicit discussions of technical debt remain limited.
Thus, we conduct a systematic Multivocal Literature Review (MLR) on technical debt and its early signals in LLM-Assisted software development to support a holistic understanding of long-term trade-offs and compounding costs across the integration of code and governance of AI-based software systems.}

\subsection{Research Objective}
\textcolor{black}{To provide a structured overview of this emerging topic, we conduct a Multivocal Literature Review (MLR) that integrates evidence from both practitioner grey literature (e.g., blogs, online forums, industry reports) and academic publications. This dual perspective allows us to capture not only the practical concerns voiced by developers in real-world settings but also the analytical and methodological insights provided by research studies.
Our objective is to comprehensively examine the forms of technical debt and their associated signs with LLMs, the risks and challenges they introduce into software projects, and the strategies, tools, and metrics currently available or still missing for managing them.} To guide this investigation, we address the following research questions:
\begin{itemize}[leftmargin=*]
\item \textbf{RQ1:} \textit{What forms of technical debt are introduced by integrating LLM-generated code into software projects?}
\item \textbf{RQ2:} \textit{What strategies or guidelines exist to mitigate technical debt introduced by LLMs?}
\item \textbf{RQ3:} \textit{What tools or techniques currently exist to help practitioners detect or measure technical debt in LLM-generated code?}
\item \textbf{RQ4:} \textit{Are there benchmarks or datasets available to evaluate the technical debt of LLM-generated code?}
\item \textbf{RQ5:} \textit{Do existing technical debt metrics sufficiently capture the debt introduced by LLMs? If not, what new metrics or dimensions are being suggested?}
\end{itemize}

\subsection{Contributions}
Our review highlights the complex landscape of integrating LLM-generated code into software systems and its implications for technical debt. Through a systematic examination of the existing literature, we analyze how 13 types of technical debt (as described in Alves et al.’s taxonomy~\cite{6974882}) manifest in the context of LLMs, identify novel forms of debt unique to the usage of LLMs, and map the strategies, tools, and metrics discussed for their mitigation. In doing so, we also surface research gaps and point toward future directions that the community should consider.
The specific contributions of this literature review are as follows:
\begin{itemize}[leftmargin=*]
\item We systematically analyze \textcolor{black}{31 formal and 73 grey sources}, providing the first multivocal perspective on technical debt in LLM-generated code.
\item We examine how different types of traditional technical debt manifest in LLM-generated code and show that traditional debts such as \textit{code}, \textit{design}, and \textit{documentation} debt are the most frequently observed in LLM-generated code, often arising from issues such as poor testing and lack of governance.
\item We highlight emerging, LLM-specific forms of debt (e.g., prompt debt, provenance debt) that go beyond the traditional definitions of technical debt.
\item We synthesize reported strategies (e.g., human-in-the-loop frameworks) and tools (e.g., SonarQube, Snyk) aimed at mitigating technical debt in LLM-augmented software development.
\item We highlight key gaps, including the lack of standardized benchmarks, the absence of LLM-specific metrics (e.g., semantic consistency of LLM outputs), and the limited availability and scope of tools for detecting technical debt induced by LLMs, thereby outlining concrete opportunities for future research.
\end{itemize}
All of our assessments, along with additional materials, are available in our supplementary materials~\footnote{https://figshare.com/s/6d94133d27a880729d6b}.

\subsection{Organization of the Paper}
Section~\ref{sec:2} provides a background on technical debt studies in software engineering. Section~\ref{sec:3.1} expands on the study’s objectives and research questions. Section~\ref{sec:3.2} details the search strategy and source selection process, followed by Section~\ref{sec:3.3}, which describes our data extraction and synthesis procedures. 
\textcolor{black}{Section~\ref{sec:4} presents the findings of our multivocal literature review by presenting evidence extracted directly from the analyzed sources for each research question, and Section~\ref{sec:5} builds on these results by synthesizing the findings, discussing their implications for researchers and practitioners, identifying open research gaps, and outlining directions for future work.} Section~\ref{sec:6} provides threats to validity, and finally, section~\ref{sec:7} presents the conclusions.
\section{Related Work}\label{sec:2}
\subsection{\textcolor{black}{TD in Traditional Software Systems}}
Technical debt in software has been widely studied through numerous secondary studies and mapping studies. Prior research has primarily focused on traditional software development settings, examining debt types such as code, design, and architectural debt, and investigating dimensions such as measurement, prioritization, tool support, and management strategies.

Besker et al.~\cite{BESKER20181} synthesized knowledge on architectural debt and proposed a descriptive model to classify and analyze architectural debt, while Sousa et al.~\cite{10.1145/3613372.3613399} and Koulla Moulla et al.~\cite{Koulla} reviewed methods and tools for the identification and measurement of architectural debt. Other works studies examined approaches for quantifying code, design, and architecture technical debt~\cite{10.1145/3675393}, techniques for prioritizing technical debt~\cite{10.1145/3387906.3388630, LENARDUZZI2021110827}, and requirements-related debt~\cite{perera2023quantifying, 10.1016/j.jss.2022.111483}. Several studies also investigated technical debt in specific contexts, such as agile development~\cite{Behutiye_2017}, microservices~\cite{Villa}, continuous engineering~\cite{lunde_continuous_2020}, and systems engineering~\cite{Kleinwaks}. Collectively, these reviews have produced taxonomies, roadmaps, and decision-making models that strengthen understanding of technical debt in traditional software projects.

Over the past decade, significant advances have been made in detecting design debt at the architectural and design levels. Prominent tools include Designite~\cite{designite}, DV8~\cite{mo:tse2019,xiao2016archdebt_icse2016,dv8}, Structure101~\cite{structure101} (recently acquired by SonarQube), Archinaut~\cite{cervantes:td2020}, and Arcan~\cite{Fontana:arcan2017,Arcelli16}. CodeScene~\cite{codescene:codescene} represents a novel approach by identifying technical debt based on revision history analysis. In addition, a wide range of other tools detect technical debt in various forms, including SonarGraph~\cite{sonargraph}, CodeInspector~\cite{codeinspector}, CodeMRI~\cite{codemri}, SQuORE~\cite{squore}, SymfonyInsight~\cite{symfony}, inFusion~\cite{paiva_evaluation_2017}, JDeodorant~\cite{paiva_evaluation_2017}, PMD~\cite{pmd}, Checkstyle~\cite{checkstyle}, and JSpIRIT~\cite{jspirit}.
Fontana et al.~\cite{fontana:jot2012} were the first to conduct a comparative evaluation of different code smell detection tools. They analyzed six versions of a software system using four tools: Checkstyle, inFusion, JDeodorant, and PMD. Fernandes et al.~\cite{Fernandes:ease2016} later presented a systematic literature review covering 29 tools, each evaluated individually for two specific smells: Large Class and Long Method. Their study involved two software systems and compared the tools using three metrics: agreement, recall, and precision. Thanis et al.~\cite{paiva_evaluation_2017} further assessed the accuracy of these four tools in detecting three code smells: God Class, God Method, and Feature Envy.
\textcolor{black}{Code smells are widely recognized as indicators of potential technical debt, often signaling deeper issues that may accumulate over time if left unaddressed. Therefore, research on code smell detection is closely related to technical debt management and provides important context for understanding how early quality issues can evolve into long-term liabilities.}
More recent research includes the work of Avgeriou et al.~\cite{avgeriou:ieee-software2021}, which compared various technical debt detection tools, and the study by Biazotto et al.~\cite{biazotto:ist2024} that examined the state of the art in automated technical debt detection. Similarly, Amanatidis et al.~\cite{amanatidis:esem2020evaluating} and Lefever~\cite{lefever:icse-seip2021} investigated the accuracy and agreement among multiple technical debt detection tools, revealing a persistent lack of consensus on what constitutes real technical debt.
To the best of our knowledge, most of these techniques and tools, particularly those operating at the architecture or design level, have not yet been applied to software systems generated by large language models (LLMs).

\subsection{\textcolor{black}{TD in AI/ML Systems}}
\textcolor{black}{More recently, studies have begun to explore technical debt in AI and ML systems, primarily focusing on the development and operation of the models themselves rather than on the downstream integration of their outputs in software projects.}
Bogner et al.~\cite{9463054} identified new debt types in AI-based systems, such as data, model, configuration, and ethics debt, showing how these differ from conventional forms by arising directly from training data quality, model lifecycle management, and ethical risks in deployment. Albuquerque et al.~\cite{10.1109/TSE.2022.3214764} reviewed the use of intelligent techniques (ML, AI, data mining) for managing technical debt, highlighting how AI itself is increasingly being applied as a tool for debt detection and remediation. Saeeda et al.~\cite{eInformatica2024Art01} broaden the scope by examining non-technical debts (process, people, social, organizational), emphasizing that AI adoption not only creates new technical challenges but also reshapes collaboration, organizational practices, and cultural norms around debt management.
Moreschini et al.~\cite{MORESCHINI2026112599} examine the evolution of technical debt from DevOps to the most recent generative AI stage, finding that Machine Learning Operations (MLOps) practices are increasingly recognized as essential for managing data-related, infrastructure and pipeline-related technical debt, especially in relation to dynamic data dependencies and model retraining. They also emphasize that AI-generated artifacts and automated pipelines introduce new governance and maintainability challenges. Together, these signal a gradual shift in the literature from studying technical debt purely in traditional software systems toward capturing the broader, system-level, and socio-technical debts that emerge in AI and LLM-enhanced development.

\textcolor{black}{Despite this growing body of work on technical debt, most studies concentrate on classical technical debt types or ML-specific issues, such as model versioning, concerning the model deployment aspect of organizations that build and maintain AI systems. Very few studies explicitly examine the role of LLMs or the debts that arise when developers use them in workflows, or when LLM outputs are integrated into software projects.
This review is among the first to bridge that gap by providing a multivocal and up-to-date perspective, drawing on both academic and practitioner sources, focusing on technical debt that emerges when LLMs and their outputs are integrated into software development. We show how LLMs not only amplify traditional forms of debt but also give rise to entirely new categories that demand fresh attention in software engineering research and practice, and what strategies we can use to tackle these challenges in the future.}
\section{Methodology}
\begin{figure*}[ht!]
    \centering
    \includegraphics[width=0.85\textwidth]{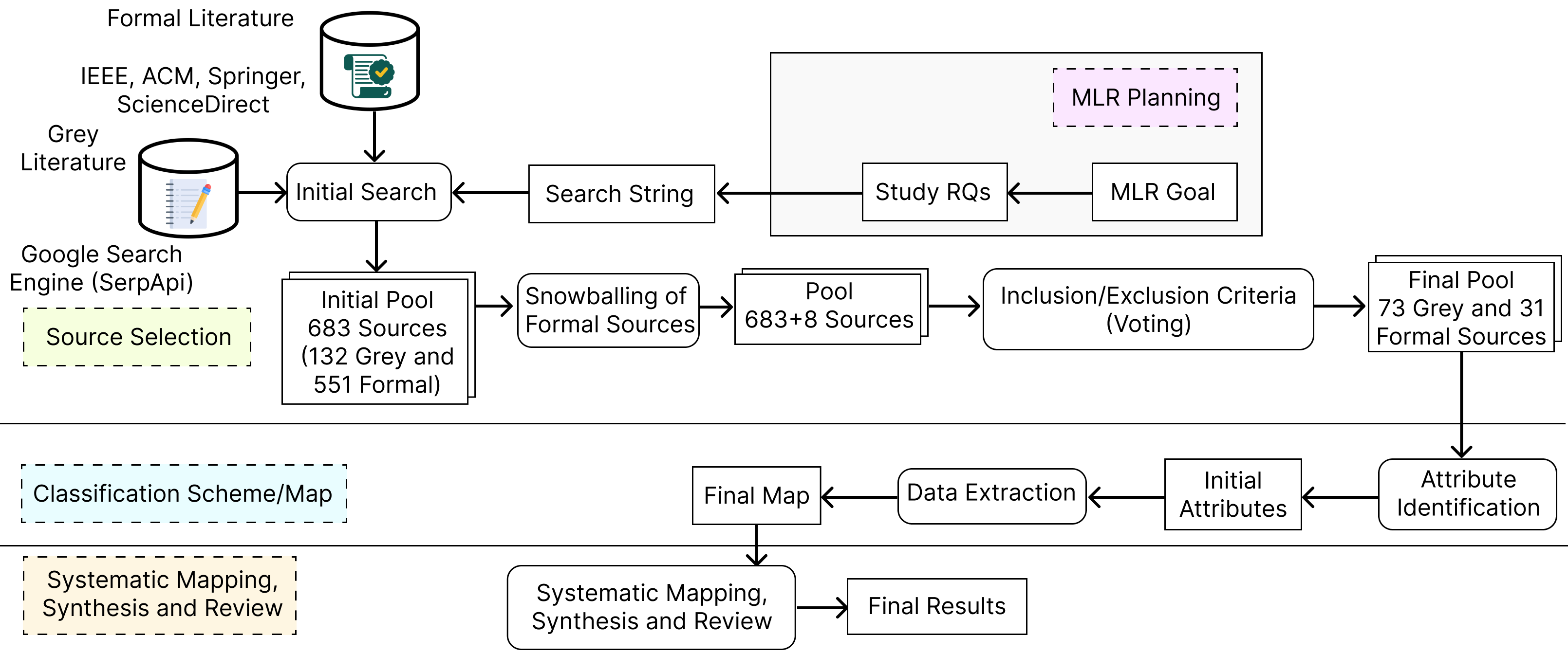}
    \caption{\textcolor{black}{An Overview of our Literature Review Process.}}
    \label{fig:mlr_design}
\end{figure*}

Our MLR process is summarized in Figure~\ref{fig:mlr_design}. The process begins with \textit{MLR Planning} phase, where we define the study goals and research questions. Next, in the \textit{Source Selection} phase, we first search formal literature and grey literature using a specific search string on a variety of data sources. We then apply inclusion and exclusion criteria through independent voting and discussion, resulting in a final pool of selected sources.
Next, we move on to the \textit{Classification Scheme/Map} phase, where we conduct attribute identification and data extraction, capturing both metadata (e.g., authors, year, source type) and study-specific fields (e.g., debt types, strategies, tools, metrics). These attributes are refined through iterative coding and reliability checks. Finally, in the \textit{Systematic Mapping, Synthesis and Review} phase, we organize and compare findings from grey and formal sources, creating a classification map of debt types, strategies, tools, and metrics, and synthesizing results to answer the research questions.
To conduct our literature review, we follow the widely adopted MLR guidelines for SE proposed by Garousi et al.~\cite{GAROUSI2019101}, while also adhering to the SE Guidelines for Reporting Secondary Studies (SEGRESS) by Kitchenham et al.~\cite{9772383} built on PRISMA. \textcolor{black}{In addition, we draw inspiration from recent thematically related review studies~\cite{GAROUSI201852, 10.1145/3661167.3661236} in terms of search strategy design, grey literature selection, quality assessment, and synthesis of practitioner and academic perspectives.}

\subsection{Goal and Research Questions}\label{sec:3.1}
The goal of this study is to systematically map, review, and synthesize the current state of both academic research and practitioner perspectives on technical debt in LLM-assisted software development. We aim to uncover emerging trends, classify reported issues, and identify directions for future research, viewed from the standpoint of both researchers and practitioners.
Technical debt has long been a critical topic in software engineering, even before the rise of AI and LLM-based tools~\cite{10.1016/j.jss.2012.12.052, 10.1016/j.jss.2014.12.027}. As these tools become increasingly integrated into software development workflows, they offer benefits such as automation and reduced manual effort~\cite{10.1145/3695988}. It is also important to examine their potential long-term effects, particularly in the form of hidden or unintentional technical debt. Our study was motivated by recurring themes in blogs and interviews suggesting that \textit{``LLMs are enabling the rapid creation of technical debt at scale''}~\cite{John_Crickett_post}. Technical debt often takes time to surface within a codebase~\cite{10.1016/j.jss.2014.12.027, 10.1016/j.jss.2012.12.052, 10.1145/2786805.2786848}. Given that LLM-assisted code generation has gained more traction since 2021~\cite{10.1145/3747588}, it is important to understand the effects of increased adoption of these models on long-term software maintainability. 

One of the key aspects of an MLR is the justification for including grey literature. Garousi et al.~\cite{GAROUSI2019101} propose a set of seven guiding questions to assess whether a grey source should be included in an MLR (see Table~\ref{tab:gl-inclusion}). According to this framework, one ``yes'' answer to any of the questions supports its inclusion. In our case, while there is an emerging body of formal research on technical debt in the context of LLMs, it takes time for such work to reach publication. Grey literature provides timely, real-world perspectives that complement the gaps in academic studies. Given the increasing adoption of LLMs in both industry and research and their growing influence on software engineering practices~\cite{10.1145/3695988}, it is critical to conduct a comprehensive multivocal review. The inclusion of grey literature in our study is therefore essential to ensure that emerging practitioner concerns are represented and can inform future academic work.

\begin{table*}[htbp]
\caption{Questions to Decide whether to Include the Grey Literature in our Study.}
\small
\begin{tabular}{@{}p{0.5em} p{9.5cm} >{\centering\arraybackslash}p{2.2cm} >{\centering\arraybackslash}p{2cm}@{}}
\toprule
\# & Question & Possible Answers & MLR-TechDebt \\
\midrule
1 & Is the subject ``complex'' and not solvable by considering only the formal literature? & Yes/No & Yes \\
2 & Is there a lack of volume or quality of evidence, or a lack of consensus on outcome measurement in the formal literature? & Yes/No & No \\
3 & Is the contextual information important to the subject under study? & Yes/No & Yes \\
4 & Is it the goal to validate or corroborate scientific outcomes with practical experiences? & Yes/No & Yes \\
5 & Is it the goal to challenge assumptions or falsify results from practice using academic research or vice versa? & Yes/No & No \\
6 & Would a synthesis of insights and evidence from the industrial and academic community be useful to one or even both communities? & Yes/No & Yes \\
7 & Is there a large volume of practitioner sources indicating high practitioner interest in a topic? & Yes/No & Yes \\
\bottomrule
\end{tabular}
\vspace{1mm}
\textit{Note: One or more ``yes'' responses suggest inclusion of grey source.}
\label{tab:gl-inclusion}
\end{table*}

Based on the goals and motivations outlined above, we classify and analyze sources from both grey and formal literature to understand the discourse on technical debt from LLM-generated code and how widely it is being discussed. We investigate the following research questions to guide our study, each formulated to be concrete, answerable, and aligned with the dual perspectives of both researchers and practitioners:
\begin{itemize}[leftmargin=*]
\item \textbf{RQ1:} \textit{What forms of technical debt are introduced by integrating LLM-generated code into software projects?}
To categorize the types of technical debt, we apply the taxonomy by Alves et al.~\cite{6974882}, which includes 13 types: \textit{Architecture Debt, Build Debt, Code Debt, Defect Debt, Design Debt, Documentation Debt, Infrastructure Debt, People Debt, Process Debt, Requirement Debt, Service Debt, Test Automation Debt,} and \textit{Test Debt} (the definitions are provided in Table~\ref{tab:debt-types}). This helps identify which aspects of software are most impacted by LLM-generated code. We also examine to identify new or emerging forms of debt that may be unique to LLM usage.
\item \textbf{RQ2:} \textit{What strategies or guidelines exist to mitigate technical debt introduced by LLMs?}
We examine whether the sources provide practical suggestions, heuristics, or formal strategies for minimizing or addressing the technical debt caused by LLM-assisted code generation.
\item \textbf{RQ3:} \textit{What tools or techniques currently exist to help practitioners detect or measure technical debt in LLM-generated code?}
This question explores existing tools, such as linters, static analysis tools, or AI-specific detectors that the literature uses or recommends to identify and measure technical debt and its early signs in LLM-generated code.
\item \textbf{RQ4:} \textit{Are there benchmarks or datasets available to evaluate the technical debt of LLM-generated code?}
We investigate whether any datasets, curated repositories, or challenge sets are available that can support systematic evaluation of LLM-induced technical debt.
\item \textbf{RQ5:} \textcolor{black}{\textit{Are there existing technical debt metrics that sufficiently capture the debt introduced by LLMs? If not, what new metrics or dimensions are being suggested?}
We investigate if there are standard metrics (e.g., those used by tools like SonarQube) that are adequate to identify technical debts introduced by LLMs, or if the literature proposes new metrics or extensions to better capture the characteristics of LLM-generated code.}
\end{itemize}

\begin{table}[htbp]
\caption{Quality Assessment Checklist of Grey Literature Sources.}
\small
\centering
\begin{tabular}{@{}p{2cm} p{12cm}@{}}
\toprule
\textbf{Criteria} & \textbf{Questions} \\
\midrule
\textbf{Authority} &
Is the author associated with a reputable organization? \newline
Is the publishing organization reputable? \newline
Has the author published other work in the field? \newline
Does the author have expertise in the area? \\
\addlinespace[0.5em]

\textbf{Methodology} &
Does the source have a clearly stated aim or brief? \newline
Does the source have a clearly stated methodology? \newline
Is the source supported by authoritative, documented references or credible sources? \newline
Does it specify how technical debt (maintainability, code quality, code smell, refactoring code written by AI) was identified or measured? \newline
Does it reference any code, data, or models shared? \newline
Are any assumptions/bias clearly stated?\\
\addlinespace[0.5em]

\textbf{Objectivity} &
Are the conclusions supported by data? \newline
Is the statement transparent about potential biases? \\
\addlinespace[0.5em]

\textbf{Date} &
Does the item have a clearly stated date? \newline
How current is the piece? \newline
Does it reference recent models or frameworks? \\
\addlinespace[0.5em]

\textbf{Significance} &
Does it enrich or add something new to the research? \newline
Does it strengthen or refute a current position (takeaways)? \newline
Does it directly address technical debt? \newline
Are failure modes, scaling issues, or long-term maintenance challenges described? \newline
Are claims supported with experiments, metrics, or real-world cases? \\
\addlinespace[0.5em]

\textbf{Impact} &
Does it offer actionable insights or tools? \\
\addlinespace[0.5em]

\textbf{Outlet Type} &
1st tier source (High outlet): government report, white papers, company/technical blogs and reports from labs \newline
2nd tier source (Moderate outlet): conference slides, news articles, developer forums and platforms \newline
3rd tier source (Low outlet): personal blogs, tweets, posts \\
\bottomrule
\end{tabular}
\label{tab:gl-quality}
\end{table}

\subsection{Search and Source Selection}\label{sec:3.2}
\textcolor{black}{Our search was conducted in January 2026, and therefore, our study includes all sources available up to January 2026.
To determine the most appropriate search query, we conducted multiple rounds of exploratory search refinement and collaborative discussions to ensure comprehensive coverage of both grey and formal literature. In the initial phase, we examined individual keywords and keyword combinations (e.g., ``technical debt” AND ``LLM”, or ``AI-generated code”) to assess the relevance and comprehensiveness of the returned results. This step allowed us to observe which terms were more effective in surfacing studies directly related to our research questions and which terms introduced excessive noise. After these iterative exploratory investigations, we aggregated the terms into a query string. Through additional iterations, we refined the logical structure of the query, combining related terms under grouped operators. The final version of the query string, presented below, consistently yielded the most relevant and comprehensive set of results across databases and search engines.}
\rqbox{\textit{(``technical debt'' OR ``code debt'' OR ``design debt'' OR ``documentation debt'' OR ``infrastructure debt'' OR ``architecture debt'' OR ``test debt'' OR ``defect debt'' OR ``defect debt'' OR ``build debt'' OR ``requirement debt'' OR ``service debt'' OR ``test automation debt'' OR ``process debt'') AND (``large language models" OR ``AI-generated code" OR ``LLMs" OR ``LLM" OR ``ChatGPT" OR ``Copilot" OR ``AI code" OR ``AI assistance")}}
\textcolor{black}{We initially only used the term ``technical debt'' in our search string, but this risked missing studies that discuss specific debt types without explicitly using the umbrella term. To mitigate this limitation, we expanded the search string to include the 13 well-established technical debt categories defined by Alves et al.~\cite{6974882}, which allowed us to capture a broader and more relevant set of sources.}

\subsubsection{Grey Literature}
To identify relevant grey literature sources discussing technical debt in the context of LLMs, we used Google as our primary search engine. Prior multivocal literature reviews in software engineering have shown that Google is effective in retrieving both highly relevant and credible grey literature, due to its robust PageRank algorithm~\cite{GAROUSI201852, GAROUSI2019101}. To this end, we leveraged SerpAPI\footnote{https://serpapi.com/}, a Google Search API that returns structured JSON results, allowing us to collect and analyze search results efficiently.
\textcolor{black}{This tool also removes all the cookies and potential bias in the retrieved links, which makes it a better choice over direct engine search by individuals.}

\textcolor{black}{Our search query returned approximately 375,000 results. However, based on our observations, the most relevant and high-quality sources were concentrated within the first 8-10 pages of search results. Beyond this range, the results became increasingly noisy, often consisting of unrelated content and broad topics of LLMs and AI, tweets, or promotional videos.
Following the guidelines by Garousi et al.~\cite{GAROUSI2019101}, which recommend capping grey literature searches once relevance saturates, we limited the sources to the first 10 pages. After removing duplicate entries and filtering out formal academic papers that appeared in the Google search results, we identified 132 unique grey literature sources relevant to our study.}

To assess the credibility and quality of these sources, we followed the evaluation framework proposed by Garousi et al.~\cite{GAROUSI2019101}, adapting it to our focus on technical debt in LLM-generated code. As shown in Table~\ref{tab:gl-quality}, our quality assessment includes 22 criteria across 7 broad categories (e.g., \textit{Authority}, \textit{Methodology}), covering aspects such as source authority and methodological clarity. 
The 132 sources were divided between two authors, who independently assessed the quality of each source. Specifically, for each source, we applied all 22 criteria: 1 (meets the criterion) or 0 (does not meet the criterion). For the \textit{Outlet Type} criterion, a partial score of 0.5 was allowed to account for second-tier platforms. 
\textcolor{black}{After the initial assessments, the two authors cross-checked the scored sources to ensure consistency in interpretation and application of the criteria. Any discrepancies were discussed and resolved through iterative consensus-based review, resulting in a final agreed-upon assessment for each source. The reported agreement of 1.0 (100\%) therefore reflects consensus after reconciliation, rather than an initial inter-rater reliability statistic. To further strengthen the rigor of the process, a third author subsequently reviewed all annotations for consistency.}

\begin{figure*}[ht!]
    \centering
    \includegraphics[width=0.55\textwidth]{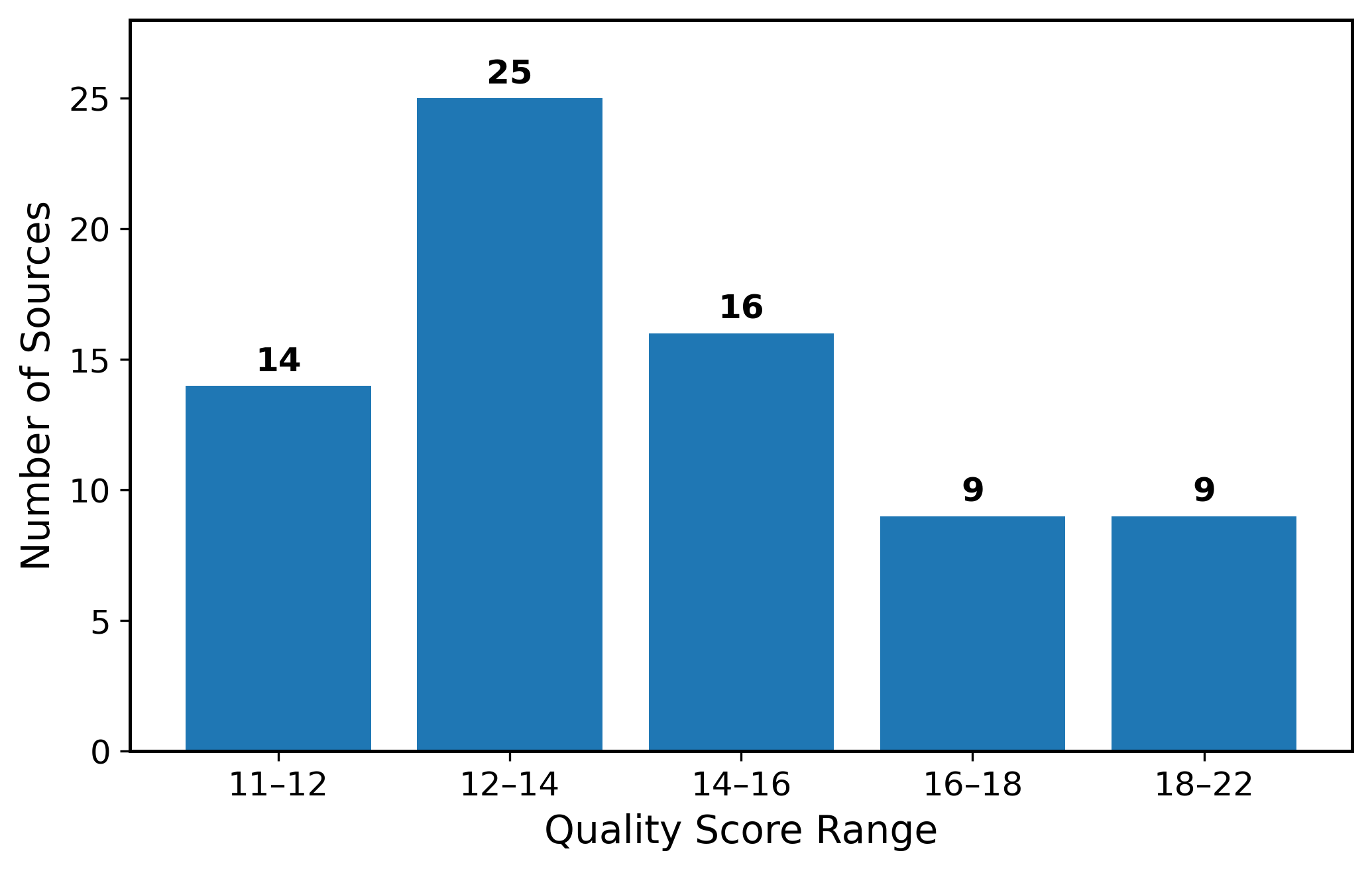}
    \caption{\textcolor{black}{Distribution of Quality Assessment Scores for Selected Grey Literature Sources.}}
    \label{fig:grey_scores}
\end{figure*}

The maximum possible score per source was 22. Following Garousi et al.’s~\cite{GAROUSI2019101} recommendation, any source scoring above 11 (i.e., at least 50\%) was deemed to have sufficient quality for inclusion. \textcolor{black}{As a result, we finalized a pool of 73 credible grey literature sources to include in our study.}
\textcolor{black}{Figure~\ref{fig:grey_scores} shows the distribution of quality assessment scores for these grey literature sources. The scores are centered around a mean of 14.4 (SD = 2.52), with a 75th percentile of 15.5, indicating that the majority of included sources meet a moderate to high quality threshold.}

\textcolor{black}{To further assess the quality and reliability of our 73 sources}, we follow the ``shades of grey'' classification model~\cite{GAROUSI2019101}, to systematically categorize them based on the outlet type, expertise of the author or organization, and degree of editorial control. This model maps sources along two dimensions of expertise and outlet control to determine their tier (1st, 2nd, or 3rd), helping assess the quality and reliability of sources.
We group our sources into four categories: \textit{Company Blogs}, \textit{Community and Developer Platforms}, \textit{Personal Blogs}, and \textit{Media Outlets}.

\begin{table}[h]
\caption{\textcolor{black}{Grey Literature Sources by Tier and Type.}}
\small
\centering
\begin{tabular}{@{}p{0.5cm}p{3cm}p{7.5cm}p{1cm}@{}}
\toprule
\textbf{Tier} & \textbf{Type} & \textbf{Sources} & \textbf{Count}\\
\midrule
1 & Company Blogs & \cite{B4, B5, B8, B9, B21, B23, B38, B41, B83, B84, B45, B51, B52, B53, B55, B60, B61, B77, B27, B28, B68, c1, c3, c4, c5, c6, c8, c10, c11, c12, c13, c14, c15, c16, c18, c19, c21, c25} & 38\\
\midrule
\multirow{2}{*}{2} & Developer Platforms & \cite{B47, B49, B55, B54.1, B67, B76, B80, B26, B30, B1, B7, B22, B81, B88, c2, c7, c9, c24} & 18 \\
   & Media Outlets & \cite{B64, B20, B3.1, B6, B10, B17, B19, B34, B36, B37, B78, c20, c22, c23} & 14\\
\midrule
3 & Personal Blogs & \cite{B3, B74, c17} & 3\\ \hline
\textbf{Total} & & & 73 \\
\bottomrule
\end{tabular}
\label{tab:grey_sources}
\end{table}

\textcolor{black}{\textit{Company Blogs} make the majority of our sources, with 38 sources.} These blogs are published by established software engineering companies, AI startups, or enterprise tool providers such as IBM, Sonar, and Salesforce. Given the clear authorship and organizational control, these sources fall into the 1st tier of grey literature, characterized by high domain expertise and moderate-to-strong editorial oversight.
\textcolor{black}{\textit{Community and Developer Platforms}, such as Medium, Hacker News, and Reddit, contribute 18 sources.} These platforms offer practitioner-driven content and valuable real-world perspectives. However, they often vary in terms of authorship transparency and editorial control, with some content being anonymous or minimally moderated. As a result, these sources fall into the 2nd tier, where moderate expertise and limited outlet control are typical.
\textcolor{black}{\textit{Media Outlets}, including sources like NewsStack and Visual Studio Magazine, account for 14 sources.} These are typically tech journalism websites, industry news platforms, or blogs affiliated with academic and professional organizations. They are generally well-edited and authored by domain-savvy writers or journalists, placing them within the 2nd tier, due to their moderate outlet control and reasonable credibility.
\textcolor{black}{Finally, \textit{Personal Blogs} such as Substack and LinkedIn posts written by independent developers, consultants, or individual practitioners make up 3 of our sources.} These sources are placed in the 3rd tier of grey literature due to their informal tone, lack of structured editorial review, and varying levels of author expertise. While they may offer early signals and novel perspectives, we interpret their content with caution during synthesis.

Therefore, our grey literature corpus, as shown in Table~\ref{tab:grey_sources}, includes materials from a mix of businesses and corporations (e.g., Salesforce), academic institutions (e.g., Johns Hopkins), media outlets, and freelance individuals.

\begin{table}[htbp]
\caption{\textcolor{black}{Number of Retrieved Papers From Each Source.}}
\small
\centering
\begin{tabular}{@{}p{4.5cm}r@{}}
\toprule
\textbf{Source} & \textbf{Number of Retrieved Papers} \\
\midrule
IEEE Xplore Digital Library & 17 \\
ACM Digital Library & 258 \\
Springer Link Online Library & 182 \\
Elsevier ScienceDirect & 94 \\
\midrule
\textbf{Total} & \textbf{551} \\
\bottomrule
\end{tabular}
\label{tab:retrieved_papers}
\end{table}

\subsubsection{Formal Literature}\label{sec:3.b.2}
\textcolor{black}{To identify relevant literature, we began by using Google Scholar and DBLP with the same search string applied in our grey literature search. However, the results were found to be noisy, consisting of personal documents hosted on individual websites or repositories. As a result, we decided to use more structured and widely accepted academic databases, following recommendations from established  guidelines~\cite{9772383}. Specifically, we searched within the \textit{ACM Digital Library}, \textit{IEEE Xplore}, \textit{Springer Link}, and \textit{Elsevier ScienceDirect}, which are frequently used in systematic and multivocal literature reviews in software engineering.}
\textcolor{black}{The number of retrieved papers from each source is shown in Table~\ref{tab:retrieved_papers}. In total, we retrieved 551 papers.} To determine which papers to include in our study, we applied a predefined set of inclusion and exclusion criteria (detailed in Table~\ref{tab:inclusion_exclusion}), developed following SEGRESS~\cite{9772383} for transparent and reproducible review processes. All included papers must be peer-reviewed publications from conferences, workshops, or journals, be related to software engineering, and explicitly discuss technical debt in the context of LLM-generated or AI-assisted code.
\textcolor{black}{To avoid ambiguity and instability in the findings of academic papers, we restrict inclusion to peer-reviewed publications because results in preprints may change substantially over time.}

\begin{table}[h]
\caption{Inclusion and Exclusion Criteria for Formal Literature.}
\centering
\small
\begin{tabular}{@{}p{1.2cm}p{12cm}@{}}
\toprule
\multicolumn{2}{@{}l}{\textbf{Inclusion Criteria}} \\
\midrule
IC1 & The paper is peer-reviewed and published in a recognized conference, journal, or workshop \\
IC2 & The paper must be available in one of the online libraries mentioned in Section~\ref{sec:3.b.2}. \\
IC3 & The paper explicitly discusses technical debt in the context of LLMs and AI-generated or AI-assisted code. \\
\midrule
\multicolumn{2}{@{}l}{\textbf{Exclusion Criteria}} \\
\midrule
EC1 & The paper is not written in English, the primary language used for technical and academic research in software engineering. \\
EC2 & The paper focuses on the functional quality or correctness of LLM-generated code but does not directly assess or discuss technical debt, and it is not the main focus of the paper. \\
\textcolor{black}{EC3} & \textcolor{black}{The paper focuses exclusively on the use of LLMs for detecting or mitigating existing technical debt, without examining how LLM-generated outputs introduce, amplify, or propagate technical debt in software artifacts.} \\
EC4 & The paper either serves as an extension of a previously accounted-for conference publication or its extended version has already been included. \\
EC5 & The paper is not a traditional research publication and instead serves as an abstract-only submission, doctoral thesis, book, poster, etc. \\
\bottomrule
\end{tabular}
\label{tab:inclusion_exclusion}
\end{table}

\textcolor{black}{The selection process was conducted manually by two authors using our set of criteria. Both authors independently reviewed the titles and abstracts of all 551 retrieved papers,} marking each for potential inclusion or exclusion based on its relevance. Disagreements or uncertainties were flagged and discussed between all authors until full consensus was reached, ensuring consistency and rigor throughout the process. \textcolor{black}{This initial screening resulted in 49 potential papers.}
\textcolor{black}{Next, we conducted a full-text review of these 49 papers, applying the inclusion and exclusion criteria more rigorously. During this step, we also extracted any relevant information necessary for answering our research questions (e.g., discussed types of technical debt, proposed tools or metrics) to make sure the papers fit the goal of our study. After this assessment, both authors agreed on the inclusion of 27 papers and excluded the remaining 22 following several rounds of discussion. The primary reasons for exclusion were that these papers were not directly related to technical debt in the context of LLMs, or they focused primarily on other aspects, such as the functional quality, correctness, or usability of LLM-generated code without addressing technical debt explicitly.
To ensure completeness and mitigate any limitations from our initial database search, we also conducted both forward and backward snowballing on the 27 selected papers. This process involved collecting all the references cited in those papers and all subsequent papers that cited them. Through this, we retrieved 1,737 papers. The two authors both manually filtered these by title, abstract, and full text as needed, leading to the inclusion of 4 additional papers. Therefore, in total, our selection process resulted in 31 papers: 27 from the initial formal search and 4 from snowballing.}

\begin{table*}[h]
\caption{\textcolor{black}{Formal Literature Sources by Venue.}}
\small
\centering
\begin{tabular}{p{1.5cm}p{8cm}p{2cm}p{1cm}}
\toprule
\textbf{Type} & \textbf{Venue} & \textbf{Sources} & \textbf{Count} \\
\midrule
Journal & Transactions on Software Engineering and Methodology & \cite{vitale_catalog_2025, autili_engineering_2025} & 2 \\
& Journal of Systems and Software & \cite{kim_comparative_2025, MORESCHINI2026112599} & 2\\
        & Information and Software Technology & \cite{ahlgren_assisting_2025} & 1 \\
        & Computing Surveys & \cite{lambiase_motivations_2024} & 1 \\
\midrule
Conference & International Conference on Software Engineering & \cite{obrien_are_2024, velasco_how_2025, batole_llm-based_2025, 11029774} & 4 \\
           & International Conference on the Foundations of Software Engineering & \cite{pomian_em-assist_2024, obrien_23_2022, 10.1145/3696630.3730565} & 3 \\
           & International Conference on Mining Software Repositories & \cite{bifolco_llms_2025, he_speed_2026} & 2\\
           & International Symposium on Empirical Software Engineering and Measurement & \cite{novielli_continuous_2024, salah_invisible_2025} & 2\\
           & International Conference on Evaluation and Assessment in Software Engineering & \cite{porta_prompt_2025, 10.1145/3756681.3756976} & 2\\
           & Brazilian Symposium on Software Quality & \cite{santos_increasing_2024, simoes_evaluating_2024} & 2\\
           & International Conference on AI Engineering & \cite{foidl_data_2022, cynthia_identification_2025} & 2\\
           & International Conference on Automated Software Engineering & \cite{zhang_copilot---loop_2024} & 1 \\
           & International Conference on Software Analysis, Evolution and Reengineering & \cite{nunes_evaluating_2025} & 1\\
           & International Conference on Technical Debt & \cite{akgul_aligning_2025} & 1\\
           & International Conference on AI Foundation Models and Software Engineering & \cite{katzy_heap_2025} & 1\\
           & Symposium on Applied Computing & \cite{aljohani_fine-tuning_2024} & 1\\
           & International Conference on Software Architecture Companion & \cite{pandini_exploratory_2025} & 1\\
           & Workshop on Machine Learning and Systems & \cite{menshawy_navigating_2024} & 1\\ 
           & Workshop on Language Models and Programming Languages & \cite{10.1145/3759425.3763390} & 1\\\hline
\textbf{Total} & & & 31 \\
\bottomrule
\end{tabular}
\label{tab:formal_sources}
\end{table*}

\textcolor{black}{As shown in Table~\ref{tab:formal_sources}, the included formal sources span a diverse set of top-tier journals, conferences, and workshops in software engineering. The journal publications come from venues such as Transactions on Software Engineering and Methodology, Empirical Software Engineering, and Journal of Systems and Software, and on the conference side, most papers appear in leading venues like FSE, ICSE, ASE, MSR, and ESEM, along with other more specialized venues such as the International Conference on Technical Debt and Workshop on Language Models and Programming Languages.}

\subsection{Data Extraction and Synthesis}\label{sec:3.3}
To answer our study's research questions, we started by conducting a data extraction process. 
Specifically, we defined a set of data extraction forms containing targeted questions to help identify key elements for answering each research question.
Following the guidelines~\cite{GAROUSI201852, 9772383}, we collaboratively designed these forms through brainstorming and discussion.
For all sources, two authors independently applied the extraction forms to all selected sources. Each author extracted data separately to ensure consistency and completeness. A third author subsequently reviewed the results and resolved any discrepancies or conflicts across the extracted data. 
For each source, we recorded general metadata (e.g., title, authors, year, source type [grey or formal], and URL), along with study-specific fields. These included:
\begin{itemize}
    \item The type(s) of technical debt discussed
    \item Any technical debt concerns unique to LLMs
    \item Proposed strategies or guidelines for mitigating technical debt introduced by LLMs
    \item Tools mentioned for measuring or detecting technical debt in LLM-generated code
    \item Benchmarks or datasets used to evaluate such debt
    \item Any new metrics or dimensions introduced beyond existing metrics (e.g., SonarQube rules)
    \item Additional notes and observations
\end{itemize}

Among the extracted fields listed above, while the first two fields required subjective judgment and interpretation, the remaining fields primarily involved identifying and extracting relevant factual information.
For formal sources, we also followed the recommendations of Kitchenham et al.~\cite{KITCHENHAM20097} for systematic reviews by computing Cohen’s Kappa coefficient to assess inter-rater reliability and reduce the risks of bias assessment. The inter-rater agreement was strong, with a Cohen’s Kappa of 0.93 for types of technical debt discussed in sources and a range of 0.90–1.00 for the rest of the questions. Any discrepancies and conflicts were resolved to finalize the data extraction of all of our sources. As advised by Garousi et al.~\cite{GAROUSI2019101}, explicit traceability links between the extracted data and primary sources are all available in our annotation sheets in the replication package.

\begin{table*}[]
\Large
\centering
\caption{\textcolor{black}{Classification of Formal and Grey Sources Based on Technical Debt Framing.}}
\label{tab:source_evidence}
\resizebox{\textwidth}{!}{%
\begin{tabular}{|p{1.6cm}|p{5cm}|p{1.2cm}|p{15cm}|}
\hline
\textbf{Evidence} & \textbf{Criteria} & \textbf{Lit.} & \textbf{Titles and Links of Sources} \\
\hline

\multirow{2}{=}{Direct TD Evidence} &
\multirow{2}{=}{(1) The source explicitly uses the term TD or a recognized category (e.g., code, design) in its title. (2) The source frames its main discussion, including findings, risks, or recommendations in terms of TD.} &
Grey &
\href{https://www.tabnine.com/blog/how-to-avoid-vibe-coding-your-way-into-a-tsunami-of-tech-debt/}{Vibe coding into a tsunami of TD~\cite{B45}},
\href{https://news.ycombinator.com/item?id=42137527}{AI makes TD more expensive~\cite{B47}},
\href{https://www.reddit.com/r/programming/comments/1it1usc/how_ai_generated_code_accelerates_technical_debt/}{How AI generated code accelerates TD~\cite{B49}},
\href{https://blog.metamirror.io/generative-ai-get-ready-to-multiply-your-technical-debt-fd60ee7d7f4e}{Generative AI: get ready to multiply your TD~\cite{B50}},
\href{https://www.lowtouch.ai/the-hidden-debt-of-ai-generated-code/}{The Hidden Debt of AI-Generated Code~\cite{B53}},
\href{https://leaddev.com/software-quality/how-ai-generated-code-accelerates-technical-debt}{AI generated code compounds TD~\cite{B1}},
\href{https://www.salesforceben.com/will-generative-ai-increase-salesforce-technical-debt/}{Will AI increase TD~\cite{B55}},
\href{https://www.castsoftware.com/pulse/artificial-intelligence-and-technical-debt-navigating-the-new-frontier}{Navigating AI and TD~\cite{B61}},
\href{https://www.hivel.ai/blog/ai-coding-assistants-superpower-or-technical-debt-generator}{AI assistants TD generator~\cite{B28}},
\href{https://www.virtasant.com/ai-today/is-ai-bloating-your-technical-debt-what-you-need-to-know}{AI bloating TD~\cite{B17}},
\href{https://devops.com/will-the-rise-of-generative-ai-increase-technical-debt/}{Rise of AI TD~\cite{B19}},
\href{https://portkey.ai/blog/the-hidden-technical-debt-in-llm-apps}{Hidden TD in LLM apps~\cite{B22}},
\href{https://www.okoone.com/spark/technology-innovation/why-ai-generated-code-is-creating-a-technical-debt-nightmare/}{TD nightmare~\cite{B78}},
\href{https://akfpartners.com/growth-blog/ai-is-it-the-new-source-of-tech-debt}{AI as new source of TD~\cite{B84}},
\href{https://www.bairesdev.com/blog/why-ai-assistants-create-technical-debt/}{Why AI assistants create TD~\cite{B88}},
\href{https://inclusioncloud.com/insights/blog/ai-generated-code-technical-debt/}{AI-generated code TD price tag~\cite{c13}},
\href{https://www.armorcode.com/blog/your-genai-code-debt-is-coming-due-heres-what-gartner-predicts}{GenAI code debt prediction~\cite{c16}},
\href{https://www.infoq.com/news/2025/11/ai-code-technical-debt/}{AI code TD report~\cite{c6}},
\href{https://www.qodo.ai/blog/technical-debt/}{TD risks~\cite{c10}},
\href{https://www.encora.com/insights/tackling-technical-debt-with-generative-ai}{Tackling TD with AI~\cite{B83}},
\href{https://www.rtinsights.com/the-pitfalls-of-generative-ai-how-to-avoid-deepening-technical-debt/}{Pitfalls of GenAI TD~\cite{B34}},
\href{https://devops.com/?p=162901}{LLMs and TD trap~\cite{B37}},
\href{https://blog.bitsrc.io/vibe-coding-the-future-of-ai-powered-development-or-a-recipe-for-technical-debt-2fd3a0a4e8b3}{Vibe coding and TD~\cite{B26}},
\href{https://kodus.io/en/ai-generated-code-is-messing-with-your-technical-debt/}{AI-generated code messing with TD~\cite{B23}},
\href{https://cerfacs.fr/coop/hpcsoftware-codemetrics-kpis}{Impact of AI-generated code on TD~\cite{B81}},
\href{https://thebootstrappedfounder.com/why-ai-generated-code-hurts-your-exit/}{AI-generated code hurts your exit~\cite{c12}},
\href{https://www.growthaccelerationpartners.com/blog/what-is-technical-debt-in-ai-generated-codes-how-to-manage-it}{TD in AI-generated code~\cite{c8}},
\href{https://medium.com/@alex.pinheiro.soares/decoding-the-old-new-technical-debt-the-risks-of-unofficial-ai-generated-iac-6230ddabf514}{Old-new AI TD risks~\cite{c7}},
\href{https://medium.com/@remotelyintelligent/the-hidden-technical-debt-in-large-language-model-llm-systems-0581e347dbc6}{TD in LLMs~\cite{B30}},
\href{https://sandar-ali.medium.com/what-is-the-technical-debt-of-large-language-models-llms-and-how-does-it-affect-us-484efe4d9552}{TD of LLMs~\cite{c9}},
\href{https://www.productreleasenotes.com/p/the-dark-side-of-ai-prototyping-technical}{Dark side of AI prototyping: TD~\cite{c22}},
\href{https://thenewstack.io/how-to-use-self-healing-code-to-reduce-technical-debt/}{Self-healing code to reduce TD~\cite{B10}},
\href{https://www.coderabbit.ai/blog/reduce-tech-debt-ai-s-role-in-efficient-coding}{AI role in reducing debt~\cite{B27}},
\href{https://www.kyndryl.com/us/en/about-us/news/2024/10/how-ai-eliminates-tech-debt-improves-software-development}{How AI can eliminate tech debt~\cite{B60}},
\href{https://engineering.jhu.edu/news/overcoming-ethical-debt/}{Overcoming Ethical debt of AI~\cite{B64}},
\href{https://aifordevelopers.io/how-software-developers-can-use-llms-to-pay-technical-debt/}{Using LLMs to pay TD~\cite{B76}},
\href{https://www.ibm.com/think/topics/technical-debt}{What is AI TD~\cite{B77}},
\href{https://medium.com/@API4AI/reducing-technical-debt-with-ai-code-review-tools-5ec03a3c8d57}{Reducing TD with AI tools~\cite{B80}},
\href{https://matrixmarketinggroup.com/github-copilot-reducing-technical-debt/}{Copilot reducing TD~\cite{B8}},
\href{https://www.alixpartners.com/insights/102jlar/can-ai-solve-the-rising-costs-of-technical-debt/}{Can AI solve the rising costs of TD?~\cite{B9}},
\href{https://metabob.com/blog-articles/preventing-technical-debt-with-ai-code-reviews.html}{Preventing TD~\cite{B21}},
\href{https://www.semasoftware.com/blog/stack-overflow-podcast-how-sema-prevents-technical-debt-from-genai-code-2}{Preventing TD from GenAI code~\cite{B41}},
\href{https://www.oreilly.com/radar/building-ai-resistant-technical-debt/}{AI-resistant TD~\cite{c3}},
\href{https://www.aei.org/technology-and-innovation/the-hidden-danger-of-security-code-debt-why-thorough-vetting-is-crucial-before-deployment/}{Dangers of AI Security code debt~\cite{c4}},
\href{https://read.dukeupress.edu/critical-ai/article-abstract/doi/10.1215/2834703X-11205182/390857/The-Moral-Hazards-of-Technical-Debt-in-Large}{Moral hazards of TD in LLMs~\cite{c5}},
\href{https://learn.castsoftware.com/reduce-cost-and-tech-debt}{Reduce cost and tech debt of AI code~\cite{c15}},
\href{https://www.unite.ai/can-developers-embrace-vibe-coding-without-enterprise-embracing-ai-technical-debt/}{Vibe coding without enterprise TD~\cite{B36}},
\href{https://futurecio.tech/balancing-the-benefits-of-ai-with-technical-debt/}{Balancing benefits of AI with TD~\cite{B7}},
\href{https://www.cio.com/article/1313757/how-cios-navigate-generative-ai-in-the-enterprise.html}{CIOs navigating GenAI and debt~\cite{B6}},
\href{https://blar.io/blog/technical-debt-part-2-llms-ai-and-the-new-frontier}{TD and LLMs frontier~\cite{B4}},
\href{https://tomtunguz.com/hidden-technical-debt-in-ai/}{Hidden TD in AI~\cite{c17}},
\href{https://sourcery.ai/blog/chatgpt-maintainable-code}{Generating Code without Generating TD?~\cite{B52}},
\href{https://www.linkedin.com/pulse/how-writing-code-ai-tools-adds-technical-debt-tshepang-mangoejane-pwd4f}{How Writing code with AI tools adds TD~\cite{B74}},
\href{https://muuktest.com/blog/technical-debt-in-agile}{Managing Agile TD with LLMs~\cite{B38}}
\\ \cline{3-4}

& & Formal &
\href{https://blog.metu.edu.tr/ttemizel/files/2025/04/2025035965.pdf}{Aligning Data Debt with AI-Integrated Software Project~\cite{akgul_aligning_2025}},
\href{https://dl.acm.org/doi/10.1145/3540250.3549088}{23 shades of self-admitted TD~\cite{obrien_23_2022}},
\href{https://dl.acm.org/doi/10.1145/3642970.3655840}{Navigating Challenges and TD in LLMs~\cite{menshawy_navigating_2024}},
\href{https://ieeexplore.ieee.org/document/11323418}{A Vision for Understanding Ethical Debt in AI-Based Coding~\cite{salah_invisible_2025}},
\href{https://doi.org/10.1145/3696630.3730565}{Automated Technical Debt Remediation with LLMs~\cite{10.1145/3696630.3730565}},
\href{https://doi.org/10.1145/3756681.3756976}{PromptDebt: A Comprehensive Study of TD Across LLM Projects~\cite{10.1145/3756681.3756976}},
\href{https://doi.org/10.1016/j.jss.2025.112599}{The Evolution of TD from DevOps to Generative AI~\cite{MORESCHINI2026112599}}
\\ \hline

\multirow{2}{=}{Supporting TD Evidence} &
\multirow{2}{=}{(1) The source's title discusses phenomena (e.g., code smells, maintainability issues, quality degradation) commonly associated with TD. (2) It discusses signs of TD accumulation in software, and has an explicit connection to TD concepts and its categories.} &
Grey &
\href{https://www.legitsecurity.com/aspm-knowledge-base/ai-code-generation-benefits-and-risks}{AI risks and benefits~\cite{B51}},
\href{https://overcast.blog/15-ai-code-refactoring-tools-you-should-know-50cf38d26877}{AI Refactoring tools~\cite{B67}},
\href{https://cybernews.com/ai-news/ai-coding-assistants-productivity-quality-tradeoffs/}{Productivity vs quality of AI~\cite{B20}},
\href{https://www.site24x7.com/learn/how-to-improve-code-quality-using-chatgpt.html}{Improve AI code quality~\cite{B5}},
\href{https://graphite.com/blog/ai-code-review-for-ai-generated-code}{AI code review~\cite{c11}},
\href{https://codescene.com/use-cases/ai-code-quality}{AI code guardrails~\cite{c14}},
\href{https://www.infoworld.com/article/3844363/why-ai-generated-code-isnt-good-enough-and-how-it-will-get-better.html}{AI code limitations~\cite{c20}},
\href{https://www.timextender.com/blog/data-empowered-leadership/challenges-with-ai-generated-code}{Challenges of AI code~\cite{c25}},
\href{https://polcode.com/resources/blog/ai-generated-code-say-hello-to-legacy-2-0/}{AI Generated code reshaping software~\cite{c21}},
\href{https://sloanreview.mit.edu/article/the-hidden-costs-of-coding-with-generative-ai/}{Hidden costs of GenAI~\cite{c23}},
\href{https://jessewarden.com/2024/01/github-copilot-research-finds-downward-pressure-on-code-quality.html}{Copilot quality~\cite{B3}},
\href{https://learningnetwork.cisco.com/s/blogs/a0DKd00006bjy76MAA/escaping-abstraction-why-aigenerated-code-demands-more-than-just-trust}{AI abstraction risks~\cite{c1}},
\href{https://visualstudiomagazine.com/articles/2024/01/25/copilot-research.aspx}{Copilot quality report~\cite{B3.1}},
\href{https://www.sonarsource.com/solutions/ai/}{Vibe then verify AI code~\cite{B68}},
\href{https://ait.inc/tech-stuffs/how-ai-generated-code-is-reshaping-software-architecture/}{AI reshaping architecture~\cite{c18}},
\href{https://www.askflux.ai/blog/ai-generated-code-revisiting-the-iron-triangle-in-2025}{AI and iron triangle~\cite{c19}},
\href{https://adabeat.com/fp/functional-programming-and-generative-ai/}{Functional programming and AI~\cite{c24}},
\href{https://stackoverflow.blog/2024/12/19/developers-hate-documentation-ai-generated-toil-work/}{AI and Documentation~\cite{c2}},
\href{https://www.linkedin.com/posts/ritzsteytler_how-ai-generated-code-compounds-technical-share-7302615149125795840-XNsr/}{New generation of AI-augmented coding~\cite{B54.1}}
\\ \cline{3-4}

& & Formal &
\href{https://courtney-e-miller.github.io/papers/SpeedAtTheCostofQuality_TheImpactofLLMAgentAssistantonSoftwareDevelopment.pdf}{How AI Increases Short-Term Velocity and Long-Term Complexity in Open-Source Projects~\cite{he_speed_2026}},
\href{https://doi.org/10.1145/3707457}{A Catalog of Data Smells for Coding Tasks~\cite{vitale_catalog_2025}},
\href{https://ieeexplore.ieee.org/stamp/stamp.jsp?arnumber=11015021}{An Study on Architectural Smell Refactoring Using LLMs~\cite{pandini_exploratory_2025}},
\href{https://dl.acm.org/doi/10.1145/3597503.3639176}{Are Prompt Engineering and TODO Comments Friends or Foes?~\cite{obrien_are_2024}},
\href{https://www.sciencedirect.com/science/article/pii/S0950584925001715}{Assisting early-stage software with LLMs~\cite{ahlgren_assisting_2025}},
\href{https://www.sciencedirect.com/science/article/pii/S0164121225001876}{Analysis of design pattern implementation in LLMs~\cite{kim_comparative_2025}},
\href{https://doi.org/10.1145/3674805.3695393}{Continuous Quality Improvement of AI-based Systems~\cite{novielli_continuous_2024}},
\href{https://doi.org/10.1145/3663529.3663803}{Safe Automated Refactoring with LLMs~\cite{pomian_em-assist_2024}},
\href{https://dl.acm.org/doi/10.1145/3712006}{Engineering Digital Systems for Humanity~\cite{autili_engineering_2025}},
\href{https://doi.org/10.1145/3701625.3701650}{Evaluating Source Code Quality with LLMs~\cite{simoes_evaluating_2024}},
\href{https://doi.org/10.1145/3605098.3636058}{An Investigation of Test Smells in Code Generation~\cite{aljohani_fine-tuning_2024}},
\href{https://www.computer.org/csdl/proceedings-article/cain/2025/021900a261/27AS7fmDfIk}{Identification of Redundant Code Using LLMs~\cite{cynthia_identification_2025}},
\href{https://doi.org/10.1145/3701625.3701637}{Increasing Test Coverage using LLMs~\cite{santos_increasing_2024}},
\href{https://doi.org/10.1145/3704806}{Challenges for Conversational Agents in SE~\cite{lambiase_motivations_2024}},
\href{https://arxiv.org/abs/2501.12134}{Do LLMs Provide Links to Code Similar to what they Generate?~\cite{bifolco_llms_2025}},
\href{https://arxiv.org/abs/2401.14176}{Fixing Code Smells in Copilot Code using Copilot~\cite{zhang_copilot---loop_2024}},
\href{https://arxiv.org/abs/2203.10384}{Categories and detection of suspicious data in AI-based systems~\cite{foidl_data_2022}},
\href{https://ieeexplore.ieee.org/abstract/document/11052803}{A Contamination-Free Code Dataset for Evaluating LLMs~\cite{katzy_heap_2025}},
\href{https://ieeexplore.ieee.org/abstract/document/10992379}{Evaluating LLMs in Fixing Maintainability Issues~\cite{nunes_evaluating_2025}},
\href{https://arxiv.org/abs/2504.13656}{Do Prompt Patterns Affect Code Quality?~\cite{porta_prompt_2025}},
\href{https://ieeexplore.ieee.org/abstract/document/11023972}{How Propense Are LLMs at Producing Code Smells?~\cite{velasco_how_2025}},
\href{https://fraolbatole.github.io/assets/pdf/LocalizeAgent.pdf}{Agent for Code Design Issue Localization~\cite{batole_llm-based_2025}},
\href{https://doi.org/10.1145/3759425.3763390}{Vibe Coding Needs Vibe Reasoning~\cite{10.1145/3759425.3763390}},
\href{https://doi.org/10.1109/ICSE55347.2025.00245}{Evaluating Deprecated API Usage in LLMs~\cite{11029774}}
\\ \hline

\end{tabular}
}
\end{table*}

\textcolor{black}{In addition, to further ensure clarity in our data extraction, we classified all sources based on how explicitly they frame technical debt. Specifically, we organize our sources into two categories: \emph{direct technical debt evidence}, and \emph{supporting technical debt evidence}. A source is categorized as \textbf{\emph{direct TD evidence}} if it (1) explicitly uses the term technical debt or its categories (e.g., code, design, documentation) in its main title, and (2) frames its main discussion, including findings, risks, or recommendations in terms of technical debt.
A source is categorized as \textbf{\emph{supporting TD evidence}} if it (1) discusses phenomena commonly associated with technical debt (e.g., code smells, maintainability issues, quality degradation) in its title, and (2) establishes an explicit connection to technical debt concepts or indicates potential debt accumulation, even if technical debt is not the primary framing. This classification was applied consistently across all sources and is used to guide our interpretation of results. The full categorization of sources is presented in Table~\ref{tab:source_evidence}.}

In the data synthesis phase, we organized and interpreted all the extracted data from these sources to derive meaningful insights from both formal and grey literature. The first two authors collaboratively conducted a qualitative analysis, synthesizing findings and drafting the initial results.
We used an approach combining deductive coding guided by the existing taxonomy of technical debt from Alves et al.~\cite{6974882} and inductive coding, where new categories were developed iteratively based on recurring patterns in the sources~\cite{KITCHENHAM20097}. For example, deductive coding was applied when classifying mentions of technical debt types (e.g., architectural debt, documentation debt), while inductive coding was used to capture emerging concepts unique to the usage of LLMs.
All synthesized results were then reviewed and discussed by all authors during multiple meetings to ensure accuracy, completeness, and consensus.

To facilitate a comparative analysis, we present the findings from formal literature, which represent the academic perspective, and those from grey literature, which reflect the practitioner perspective. This separation allows us to highlight key differences and commonalities in how technical debt related to LLMs is perceived and addressed across academic and industry contexts.

\section{Results}\label{sec:4}

\subsection{RQ1: What forms of technical debt are introduced by integrating LLM-generated code into software projects?}

\textcolor{black}{Table \ref{tab:debt-types} summarizes the identified types of technical debt, along with their definitions and frequencies across the formal and grey literature. The Source column distinguishes between (1) 13 technical debt types drawn from the established taxonomy of Alves et al., which we used as the starting point for answering this research question, and (2) 6 additional debt types that emerged from our multivocal literature review and are not covered by the Alves et al. taxonomy.}

\begin{figure*}[ht!]
    \centering
    \includegraphics[width=0.6\textwidth]{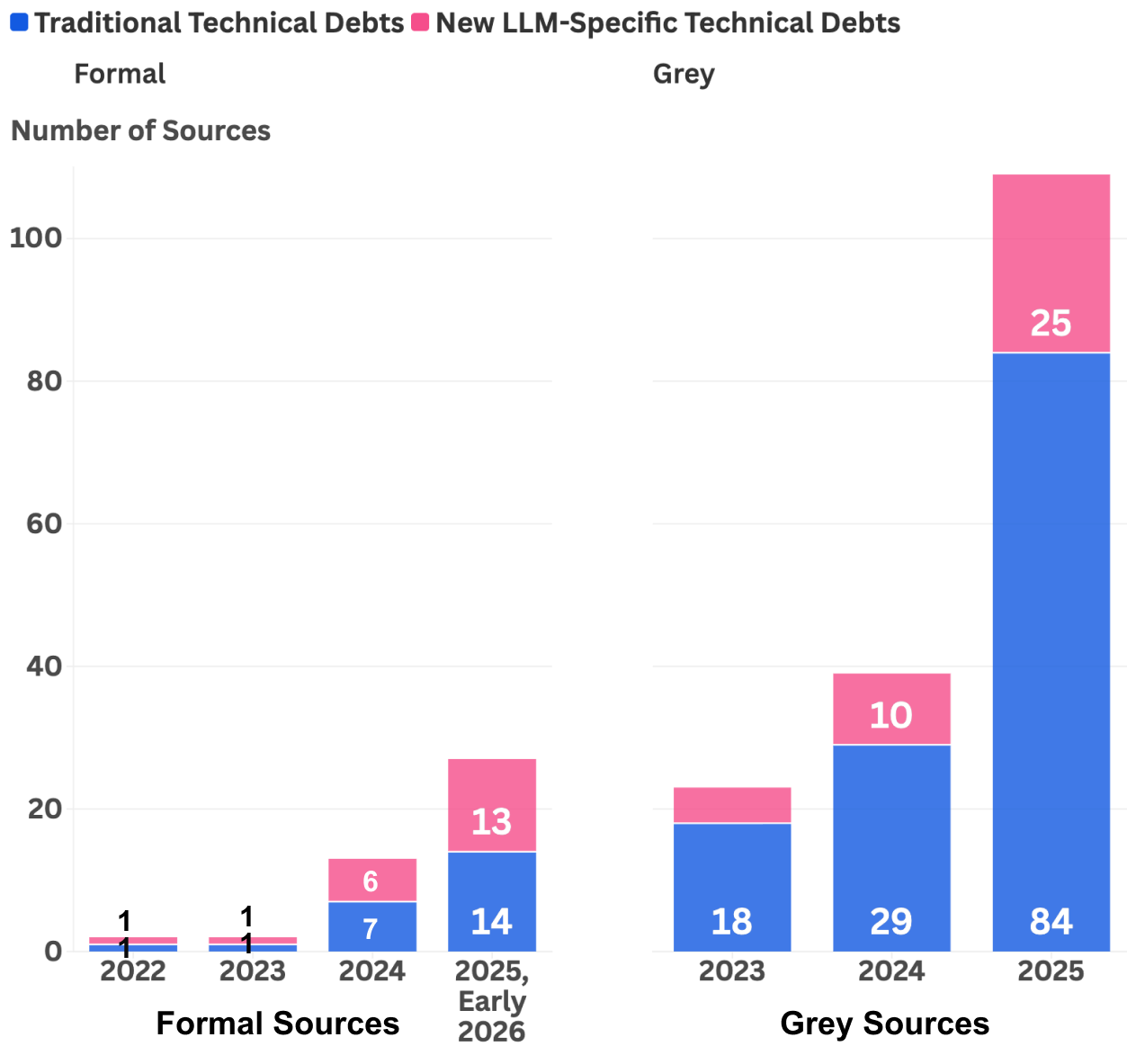}
    \caption{\textcolor{black}{The Number of Grey and Formal Sources Addressing Traditional and New Technical Debts of LLMs per Year.}}
    \label{fig:sources_per_year}
\end{figure*}

\textcolor{black}{Since 2023, there has been a growing awareness of the technical debt risks introduced by incorporating LLMs into software development workflows among practitioners. Figure~\ref{fig:sources_per_year} specifically shows the frequency of mentions in each source type by year. We observe a steady increase in both sources since 2023, reflecting growing concerns about technical debt introduced by LLM-assisted development.}

\subsubsection{Grey Literature}

\textcolor{black}{Across all the 73 sources reviewed, we identified two major categories of concerns: traditional forms of technical debt that were worsened by the use of LLMs, and emerging forms of debt unique to the use of LLMs.}

\begin{table}[h]
\caption{\textcolor{black}{Technical Debt Types with their Definitions and Frequency of Mentions Across Formal and Grey Literature. MLR Denotes Debt Types Identified in this Review.}}
\label{tab:debt-types}
\resizebox{\textwidth}{!}{%
\begin{tabular}{|l|l|p{8cm}|ll|}
\hline
\multirow{2}{*}{\textbf{Source}} & \multirow{2}{*}{\textbf{Type of Debt}} & \multirow{2}{*}{\textbf{Definition}} & \multicolumn{2}{l|}{\textbf{Frequency of Mentions}} \\ \cline{4-5} 
 &  &  & \multicolumn{1}{l|}{\textbf{Formal}} & \textbf{Grey} \\ \hline
\multirow{13}{*}{\textbf{\begin{tabular}[c]{@{}l@{}}Alves\\ et al.~\cite{6974882}\end{tabular}}} & Code Debt & Problems in source code (e.g., poor readability, bad practices) that hinder maintainability. & \multicolumn{1}{l|}{14} & 56 \\ \cline{2-5} 
 & Design Debt & Violations of good design principles (e.g., tightly coupled or overly complex classes). & \multicolumn{1}{l|}{2} & 30 \\ \cline{2-5} 
 & Documentation Debt & Incomplete, outdated, or missing documentation that hinders correct system use or evolution. & \multicolumn{1}{l|}{0} & 17 \\ \cline{2-5} 
 & Infrastructure Debt & Weaknesses in infrastructure that delay upgrades, scaling, or other development activities. & \multicolumn{1}{l|}{1} & 12 \\ \cline{2-5} 
 & Architecture Debt & Issues in system architecture (e.g., modularity, performance, robustness) that require extensive rework. & \multicolumn{1}{l|}{1} & 10 \\ \cline{2-5} 
 & Test Debt & Gaps in testing (missing cases, low coverage, flaky or unrun tests) that reduce confidence in quality. & \multicolumn{1}{l|}{2} & 6 \\ \cline{2-5} 
 & Defect Debt & Known or deferred bugs/defects that accumulate over time, making fixes harder later. & \multicolumn{1}{l|}{3} & 1 \\ \cline{2-5} 
 & Build Debt & Inefficient or poorly defined build processes that slow development and add unnecessary cost. & \multicolumn{1}{l|}{0} & 0 \\ \cline{2-5} 
 & People Debt & Skill gaps, concentrated expertise, or insufficient training that slow progress. & \multicolumn{1}{l|}{0} & 0 \\ \cline{2-5} 
 & Requirement Debt & Partially or poorly implemented requirements that fail to meet quality or performance needs. & \multicolumn{1}{l|}{0} & 0 \\ \cline{2-5} 
 & Service Debt & Suboptimal service or substitution decisions that create long-term maintenance costs. & \multicolumn{1}{l|}{0} & 0 \\ \cline{2-5} 
 & Test Automation Debt & Lack of automated tests that limit continuous integration. & \multicolumn{1}{l|}{0} & 0 \\ \cline{2-5} 
 & Process Debt & Inefficient or outdated processes no longer suited for current needs. & \multicolumn{1}{l|}{0} & 0 \\ \hline \hline

\multirow{6}{*}{\textbf{MLR}} & Governance Debt & Long-term risks from over-reliance on LLM-generated code, requiring ongoing oversight due to issues like hallucinations and non-determinism. & \multicolumn{1}{l|}{0} & 19 \\ \cline{2-5} 
 & Prompt Debt & Caused by unclear, sensitive, or undocumented prompts that reduce reproducibility and code quality. & \multicolumn{1}{l|}{8} & 5 \\ \cline{2-5} 
 & Fast-integration Debt & Risks from rapidly integrating LLM-generated outputs without proper validation, leading to unstable foundations. & \multicolumn{1}{l|}{0} & 13 \\ \cline{2-5} 
 & Data Debt & \textcolor{black}{Hidden liabilities arising from noisy, low-quality, or opaque data used in training, fine-tuning, or retrieval, which propagate flaws into generated code.} & \multicolumn{1}{l|}{9} & 0 \\ \cline{2-5} 
 & Ethical Debt & Accumulated risks of bias, fairness, or accountability failures in LLM-generated code that affect end users. & \multicolumn{1}{l|}{3} & 2 \\ \cline{2-5} 
 & Provenance Debt & Unclear ownership or missing attribution in AI-generated code, creating legal and accountability challenges. & \multicolumn{1}{l|}{1} & 1 \\ \hline
\end{tabular}%
}
\end{table}

Among the \textbf{traditional forms of technical debt (Alves et al.~\cite{6974882})}, the most frequently discussed issue is \textit{code debt} \textcolor{black}{across 56 sources}, followed by \textit{design debt} \textcolor{black}{(30 sources)} and \textit{documentation debt} \textcolor{black}{(17 sources)}.

Many industry practitioners highlight that developers often integrate LLM-generated code without fully understanding its logic or verifying its correctness, leading to duplicated or inefficient code structures and increased \textit{code debt}. As noted in Visual Studio Magazine, \textit{``The bottom line is that using Copilot is strongly correlated with mistake code being pushed to the repo''} \cite{B3.1}.
This practice increases complexity within the codebase, introduces hidden bugs, and contributes to long-term challenges in maintainability and scalability~\cite{c6, c9, c10}. As per an article from Tabnine company, \textit{``the code generated for edge cases or intricate business logic often contains errors or simply doesn’t work, requiring the human developer to spend extra time debugging or rewriting it. In effect, the AI can become more of a hindrance than a help on non-trivial coding tasks''} \cite{B45}. 
Developers frequently share concerns about the long-term maintainability of LLM-generated code. \textcolor{black}{Incorrect or illogical code is often difficult to detect and fix, especially when developers lack the expertise to spot flaws or overly trust the model's output~\cite{c24}. When these flaws are introduced unchecked, they compound existing code debt, gradually making the system harder to maintain and evolve. Additionally, as noted by Steve Jones, Executive VP of CapGemini, Generative AI generates more code in less time, which increases code complexity and ultimately drives up maintenance costs~\cite{B50}.}

\textcolor{black}{\textit{Design debt} is another concern frequently mentioned in sources that extend these maintainability challenges. Developers emphasize that AI does not understand ``your'' system~\cite{B23}, meaning that while it can generate code in seconds, such outputs are not necessarily aligned with existing software design decisions and architecture. Without proper oversight, this can lead to code that disregards the overarching system design~\cite{c7, B53}: \textit{``AI doesn’t understand your system-it just generates code based on statistical patterns. It doesn’t consider your architecture, internal conventions, or existing functionality''}~\cite{B23}.}

This potential lack of human oversight and understanding also contributes to a growing codebase where quick AI-generated solutions without any documentation accumulate complexity and \textit{documentation debt} over time, increasing the cost of refactoring and manual developer review. An article in Blar.io talks about how \textit{``developers often hesitate to modify or refactor AI-generated code because they don’t fully grasp its implications''} \cite{B4}. Two of the sources also talk about teams losing visibility into how and why certain code decisions were made. For example, the company PortkeyAI included in their blog that \textit{``Ignoring this debt can slow down iteration, hurt user experience, and make scaling painful. New team members struggle to understand the system. Security and compliance risks go unnoticed. And costs grow uncontrollably without visibility''}~\cite{B22}.
\textcolor{black}{A developer from Cisco calls this process an \textit{Abstraction Prison}, where software grows every day with AI-generated code, but it is hard to explain how it is working: \textit{``The abstraction prison isn't about the AI being wrong (though it often is). It's about the growing gap between the code that exists and the code we understand. Every time we accept AI-generated code without truly comprehending it, this is like adding another bar to your cell. Before long, you're surrounded by a codebase that runs, passes tests, maybe even ships to production, but you can't explain how or why it works''}~\cite{c1}}.

Beyond these traditional concerns, \textcolor{black}{40 sources} discuss \textbf{five emerging forms of technical debt} 
\textcolor{black}{in the context of LLM-assisted software development, which we discuss next. Importantly, our focus here is not on technical debt in AI/ML model development, data pipelines, or MLOps infrastructure. Rather, we focus on debt that arises when LLM-generated artifacts, such as code, prompts, and related outputs, are integrated into software development and must later be understood, reviewed, and maintained.}

\textit{Governance debt} (19 sources) is the most frequent LLM-assisted software development debt discussed in grey literature. 
\textcolor{black}{In our review, this debt refers specifically to the long-term oversight burden created when LLM-generated code or logic is incorporated into software systems, rather than governance challenges in training or operating AI/ML models themselves. This includes the ongoing need for oversight due to issues such as hallucinations and model non-determinism, both of which can undermine reliability and increase the cost of sustaining AI-assisted systems over time, especially in cases where code semantically appears correct but is logically flawed~\cite{B3, B4}}. 
Based on data collected by International Data Corporation (IDC), 70\% of developers using GenAI tools need to remediate as much as 40\% of that generated code~\cite{c1}.

Another commonly reported concern is \textit{fast-integration debt} (13 sources), where developers implement AI suggestions quickly, or \textit{vibe code}, without understanding their implications, leading to unstable software foundations~\cite{B45}. \textcolor{black}{This form of debt is specific to the integration of LLM outputs into software projects, where the speed of adoption can outpace developers' ability to evaluate downstream maintainability and design consequences.}
A developer describes this in a personal experience where they had to race to finish the product using AI, which resulted in bugs surfacing after weeks of implementation: \textit{``I learned this the hard way during a sprint where we were racing to add features. We had an AI assistant churning out some React components, API handlers...Our velocity metrics looked incredible...Then came the bug reports. Not immediately, that would've been too easy, right? No, these surfaced weeks later. Edge cases AI hadn't considered''~\cite{c1}}.
Therefore, this fast integration, in turn, will contribute to the need for more governance in the future.

\textcolor{black}{\textit{Prompt debt} in 5 sources is another recurring concern that is particularly salient in LLM-assisted development. It refers to unclear or undocumented prompts that degrade code quality and hinder reproducibility, as well as the constant need to determine the proper contextual information in which to add prompts. \textcolor{black}{Unlike traditional software artifacts, prompts directly shape generated implementations, yet they are often not preserved, versioned, or maintained with the same rigor as source code.} Developers note issues like prompt brittleness, where minor changes lead to drastically different outputs (\textit{``Small inconsistencies drift into large-scale architectural entropy''}~\cite{c13}), as well as ambiguous instructions and lack of prompt versioning \cite{B49, B50}.}
 
\textit{Ethical debt} refers to fairness and bias concerns that arise when LLM-generated technical debt disproportionately affects end users.
Unlike traditional technical debt, ethical debt is often invisible to developers yet deeply consequential for affected users. While the term is increasingly discussed in the context of LLM-assisted development, ethical debt itself is not new; it has long been associated with AI-based systems, particularly in relation to algorithmic fairness, prediction bias, and accountability~\cite{9463054}. \textcolor{black}{What distinguishes the LLM setting in our review is that these risks can be introduced indirectly through generated code, configurations, or design suggestions that developers incorporate into software projects, even when they are not directly building AI/ML models or pipelines, and that such risks may be amplified in this context.
The huge scale of LLM training data, which is often unknown, combined with the lack of transparency of their mostly black-box architectures, makes biases and ethical risks integrated into software harder to detect or mitigate.} As a result, issues of fairness and accountability are not only inherited but often magnified, creating a larger and less transparent burden for developers and stakeholders. Johns Hopkins' article mentions this debt as \textit{``It’s when you’re more interested in putting out a solution quickly, which is often the business driver-to be first to market. But you haven’t yet looked at your solution from an ethical perspective and you end up having a situation where there are biases or other problems within the data or AI solution that haven’t been addressed''}~\cite{B64}.

An additional form of technical debt discussed in the context of LLM-assisted software development in one source is \textit{provenance (ownership) debt}, where the responsibility and ownership for LLM-generated logic becomes unclear as code cloning becomes more common with the use of AI tools~\cite{B1}. \textcolor{black}{Here, the concern is not ownership of models or datasets, but ownership and traceability of the generated software artifacts themselves.}

\subsubsection{Formal Literature}
\textcolor{black}{A small but growing body of formal literature examines how LLMs can contribute to technical debt. Across these studies, the most frequently discussed concerns relate to code-level and architectural issues, including code quality and maintainability risks that signal the accumulation of technical debt in software.}

\textcolor{black}{14 studies discuss \textit{code debt} and quality issues with software in the era of LLM and AI-assistant usage.
One of these studies examines the causal effect of adopting Cursor, a widely used LLM agent assistant, on development velocity and software code debt. He et al.~\cite{he_speed_2026} conduct a difference-in-differences design comparing Cursor-adopting GitHub projects with a matched control group of similar GitHub projects that do not use Cursor, finding that the adoption of the AI agent leads to a statistically significant large increase in project-level development velocity, along with a substantial and persistent increase in complexity and code debt. \textcolor{black}{Specifically, static analysis warnings increase by 30\% and code complexity increases by 41\% post-adoption, all of which are symptoms of code debt.
Wang et al.~\cite{11029774} examine seven LLMs, 145 API mappings from eight popular Python libraries, and 28,125 completion prompts. They find that all evaluated LLMs face issues with deprecated API usages inside generated code due to the absence of API deprecation knowledge during model inference, leading to the accumulation of debt when integrated into the software.}}

\textcolor{black}{One of the most frequently discussed debts (in 9 studies) is \textit{data debt}. \textcolor{black}{Because this category overlaps with prior technical debt work on AI/ML systems, we use it here in a narrower sense. In our review, data debt does not refer broadly to debt in model training pipelines or MLOps infrastructure. Rather, it refers to data-related conditions that shape LLM-generated outputs and later create maintenance burdens when those outputs are incorporated into software projects.} This debt stems from the quality of data that influences LLM behavior, including training corpora, fine-tuning datasets, and retrieved contextual data, leading to downstream effects on code generation. While model training is typically performed by LLM providers, data debt is not limited to them. An increasing number of organizations deploy domain-specific or smaller language models~\cite{belcak2025smalllanguagemodelsfuture}, apply lightweight fine-tuning, or rely on retrieval-augmented pipelines where the quality of data remains critical. Therefore, the consequences of data debt also manifest at the point of use, directly affecting organizations that integrate LLM-generated code into their software systems.}
\textcolor{black}{Velasco et al.~\cite{velasco_how_2025} introduce the Propensity Smelly Code (PSC) metric to estimate the likelihood that code generated by LLMs contains smells. The accumulation of code smells is a direct symptom of a debt that developers have to pay in future, and the authors argue that these smells often stem from the fact that LLMs are trained on publicly available codebases, which themselves may contain poor design patterns or bad practices, and therefore, can be reproduced in the generated outputs of LLMs. Similarly, Vitale et al.~\cite{vitale_catalog_2025} characterize the presence of ``noise tokens'', such as file paths, non-ASCII characters, or personally identifiable information, in LLM training data as a major signal to data debt since such tokens may distract models from learning more meaningful abstractions.}

\textcolor{black}{\textit{Prompt debt} is another type of debt discussed in the formal literature. Aljohani et al.~\cite{10.1145/3756681.3756976} describe it as debt resulting from poorly structured,
inefficient, or suboptimal inputs to the LLM systems, and they find it to be the primary source of LLM-specific debt. They also find that instruction-based prompts and few-shot prompts are particularly vulnerable due to their dependence on instruction clarity and example quality, revealing the two prompt techniques that attract the most debt.} 

Another area receiving attention is \textcolor{black}{\textit{defect debt} discussed in 3 sources}.
\textcolor{black}{These studies mention self-admitted technical debt (SATD) as an important signal of defects in LLM code. Prior research has shown that SATDs often indicate underlying technical debt and may co-occur with different types of it such as defect debt in software~\cite{6976075, 10.1145/2901739.2901742}.} Building on this perspective, a study by O’Brien et al.~\cite{obrien_are_2024} evaluates how Copilot, an LLM coding tool, handles comments containing SATDs. The authors generate over 1,000 function bodies from 380 different TODO-style prompts and find that LLMs often replicate SATD verbatim, including placeholders like ``TODO: check divide by zero''. They suggest that prompt sensitivity is a real risk: vague or incomplete instructions can lead to the blind reproduction of debt in generated code. This risk is compounded by the fact that many LLMs are trained on public codebases where such SATDs already exist, potentially reinforcing and propagating poor practices learned from the training data itself. \textcolor{black}{In a related study, Zhang et al.~\cite{zhang_copilot---loop_2024} examine Copilot-generated Python code and detect eight distinct types of code smells, further demonstrating early indicators of technical debt that LLMs not only inherit from their training data but also produce outputs that exhibit similar technical debt patterns.}

\textcolor{black}{Finally, two of the sources introduce discussions around emerging forms of legal and ethical debts \textcolor{black}{in the specific context of integrating LLM-generated artifacts into software projects.} Bifolco et al.~\cite{bifolco_llms_2025} examine whether LLM-generated code includes links or citations to its sources and conclude that such tools often lack clear attribution, resulting in what they term \textit{provenance debt}. They raise important questions around code ownership and licensing, although this topic remains underexplored compared to more traditional forms of debt. \textcolor{black}{This concern is not about model ownership or dataset governance in general, but about the traceability, attribution, and licensing concerns of generated code that becomes part of a larger software system.}
Salah et al.~\cite{salah_invisible_2025} describe \textit{ethical debt} as the accumulation of unresolved ethical risks in AI-assisted software development, and emphasize the importance of its sources, manifestations, and outcomes. They specifically highlight mapping accountability across developers, AI systems, and organizations as an important factor that needs more attention in future research.}


\rqbox{LLMs introduce technical debt in two ways:  
(1) \textit{Traditional debts}, especially \textit{Code}, \textit{Design}, and \textit{Documentation debts}, worsened by hallucinations, weak testing, and limited oversight.  
(2) \textit{New debts} unique to LLMs: grey literature stresses \textit{Prompt}, \textit{Fast-integration}, \textit{Provenance}, \textit{Ethical}, and \textit{Governance debt};  
Formal literature highlights \textit{Data}, \textit{Prompt}, \textit{Provenance}, and \textit{Ethical} debts.}

\subsection{RQ2: What strategies or guidelines exist to mitigate technical debt introduced by LLMs?}
\subsubsection{Grey Literature}
Strategies offered by grey literature to mitigate technical debt can be broadly grouped into two themes: (1) Tooling and human-in-the-loop frameworks, and (2) Prompt engineering. 

A consistent recommendation across \textcolor{black}{58 out of 73} sources is to follow a human-in-the-loop model. \textcolor{black}{Developers and teams are encouraged to treat LLMs as junior developers, using their outputs as drafts that require review and refinement. An article from Cisco defines this relationship between AI tools and developers more clearly: \textit{``Treat AI-generated code like you're conducting a real code review for, say, a junior developer who's really enthusiastic but occasionally overconfident. Because that's essentially what it is''}~\cite{c1}, and an article from O'Reilly emphasizes this more by mentioning the need to ``trust, but verify'': \textit{``Trust the AI to give you a starting point, but verify that the design supports change, testability, and clarity''}~\cite{c3}.}

Other practices mentioned include applying clean code principles to LLM-generated outputs~\cite{B1}, using unit tests to validate AI-generated suggestions~\cite{B3}, and enforcing multi-layered review processes for cross-functional oversight~\cite{B3.1, B4}. Reddit users express similar views, with developers emphasizing the need to review, test, and document the logic behind AI-generated code \cite{B49}.
Many teams are also integrating tools and establishing organizational practices to monitor and enforce quality in AI-assisted development. Tools such as SonarQube are commonly used for automatic code quality checks, vulnerability detection, and compliance enforcement~\cite{B68, B51}. Others integrate static analysis tools into AI-augmented review pipelines to reduce duplication and improve maintainability \cite{B53}. \textcolor{black}{Additionally, teams use agentic LLM tools like Cursor and HopeAI to flag issues related to maintainability, security, and code standards \cite{B54.1, B26}}. \textcolor{black}{While these tools do not directly measure technical debt but they help identify early indicators and potential risks that may contribute to debt accumulation over time.}  At an organizational level, sources highlight the importance of ongoing developer training, documenting AI involvement in codebases, and prioritizing long-term maintainability over short-term productivity gains~\cite{B7, B78}.

Another set of strategies in 4 sources focuses on improving prompt quality and defining clear boundaries for AI usage. Developers recommend maintaining prompt templates to improve reproducibility and consistency~\cite{B45}. Others suggest using structured prompts like ``act as an expert developer” or providing explicit style/quality constraints to guide the model toward more maintainable outputs~\cite{B52}. Several sources also emphasize the importance of LLM usage segmentation, clearly delineating when to use fully generative AI or manual coding, to maintain control and ensure consistent quality~\cite{B4, B50}.

The remaining articles, i.e., \textcolor{black}{11 out of 73} grey literature sources, do not discuss strategies or guidelines to mitigate technical debt introduced by LLMs. 

\subsubsection{Formal Literature}
Strategies in formal literature are largely aligned with those offered by practitioners. These strategies can be grouped into three main categories: (1) Data quality and standards alignment, (2) Prompt engineering, and (3) Tooling and human-in-the-loop frameworks.

A common recommendation across \textcolor{black}{8 studies} involves improving the quality of training and input datasets before integrating them into development pipelines. \textcolor{black}{These works emphasize rigorous preprocessing steps such as labeling, deduplication, and filtering~\cite{vitale_catalog_2025, akgul_aligning_2025, aljohani_fine-tuning_2024, katzy_heap_2025}, to reduce the risk of accumulation of downstream code smells and poor-quality generations that will inevitably cause technical debts to pay for. Some papers further advocate aligning technical debt management with industry standards. Akgul et al.~\cite{akgul_aligning_2025} propose incorporating ISO/IEC 5338/12207 practices to address data debt, suggesting that aligning with established software lifecycle standards may help mitigate long-term risks and promote sustainable AI-assisted development.}

\textcolor{black}{Prompt design and contextual richness also emerged as critical factors contributing to or alleviating various forms of debt in 7 sources. Poorly constructed inputs to the LLMs and a lack of sufficient context in prompts are repeatedly cited as primary causes of code and architectural debt in LLM-generated outputs~\cite{10.1145/3756681.3756976}, and research has explored how effective prompt engineering could reduce such debt. Zhang et al.~\cite{zhang_copilot---loop_2024} demonstrate that while Copilot-generated Python code often exhibits common smells that signal risks of technical debts, the use of well-structured and specific prompts was able to mitigate up to 87.1\% of these issues. Additionally, combining better prompts with test-driven development to systematically prevent technical debt accumulation is another strategy explored in this domain~\cite{pandini_exploratory_2025}.}

Beyond inputs and data preparation, \textcolor{black}{10 sources} in the literature emphasize that LLMs are more effective when integrated into broader tool-chains or used within reflective workflows. Rather than relying on one-shot code generation, researchers recommend using LLMs alongside post-generation quality checks or prompting refinement loops. Retrieval-Augmented Generation (RAG)~\cite{pandini_exploratory_2025} and LLM-as-Judge frameworks~\cite{kim_comparative_2025, pomian_em-assist_2024, autili_engineering_2025, zhang_copilot---loop_2024} are among the most cited strategies in this domain. These methods introduce systematic self-reflection or multi-stage review into the generation process to surface and reduce debt.
Nunes et al.~\cite{nunes_evaluating_2025} underscore the need for post-generation oversight. Their study finds that while LLMs can improve code readability and address certain maintainability concerns, they often introduce new issues or fail to resolve existing ones fully. The authors advocate for fine-tuning and maintaining human oversight throughout the development lifecycle.
\textcolor{black}{Mitchell et al.~\cite{10.1145/3759425.3763390} propose Vibe Reasoning, a theoretical system throughout the vibe coding process, to auto-formalize specifications, validate against targets, deliver actionable feedback to the LLM, and allow intuitive developer influence on specifications. Their proof of concept conclusions align with a broader consensus in the literature: effective code generation with LLMs requires high-quality data, carefully designed prompts, continuous validation, and strong human-AI collaboration.}

The remaining articles, i.e., \textcolor{black}{6 out of 31} formal literature sources, do not discuss strategies or guidelines to mitigate technical debt introduced by LLMs. 

\rqbox{Top strategies suggested by both grey and formal literature for mitigating technical debt include detailed prompt design, use of code quality tools such as SonarQube, and adding a human-in-the-loop approach to refine and review LLM-generated code before deployment. Academic sources also emphasize improving training data quality to ensure more reliable outputs.}

\subsection{RQ3: What tools or techniques currently exist to help practitioners detect or measure technical debt in LLM-generated code?}
\subsubsection{Grey Literature}
\textcolor{black}{While only a handful of tools are specifically designed to detect signs and indicators of technical debt introduced by LLM-generated code, several sources highlight a mix of general-purpose, AI-specific, and emerging tools currently in use by practitioners. These can be grouped into: (1) general-purpose static analysis tools, and (2) emerging LLM-aware tools.}

\textcolor{black}{The most frequently mentioned tool across 4 sources is SonarQube, which remains a go-to for detecting early signals of technical debt such as code smells, and duplication, even though it is not explicitly tailored to LLM-generated outputs. Other commonly used tools include linters such as ESLint and Pylint, which help flag security vulnerabilities, code smells, and performance bottlenecks that can cause technical debt in software. However, like SonarQube, these tools are designed for general use and not optimized for the nuances of AI-generated code~\cite{B68, B74}.}
CAST Software is mentioned in \textcolor{black}{2 sources}, but again in the context of traditional technical debt detection rather than LLM-specific insights~\cite{B61, c15}.

A number of newer tools are mentioned for their emerging capabilities to analyze or support LLM-assisted development. Amazon CodeWhisperer is noted for its features to check LLM-generated recommendations for security issues in code~\cite{B20, B67}. GitHub Copilot, while primarily a code generation tool, is gradually introducing responsible AI features, such as flagging insecure or biased outputs and including reference citations~\cite{B8, c8},~\cite{copilot365}. For example, it uses built-in protections to block harmful content and detect prompt injections, which are attempts to manipulate the tool into producing unsafe output. \textcolor{black}{CRken and CodeScence} are highlighted as tools for automatically reviewing merge requests and flagging ``quick fixes'' that may introduce debt over time~\cite{B80, c14}. \textcolor{black}{Agents such as Swimm, HopeAI, and Graphite are being used to enhance documentation and explainability to avoid technical debt in the future \cite{B20, B80, B67, c11}}. \textcolor{black}{Snyk, DeepSource, ArmorCode, and Codacy} are cited as helpful for detecting \textcolor{black}{early indicators of technical debt such as} code smells, dependency issues, and maintainability risks in LLM-assisted workflows~\cite{B67, c16}.
Tabnine is mentioned for its support for model fine-tuning in enterprise environments, aiming to reduce inconsistency and maintain alignment with internal coding standards~\cite{B45}.

\subsubsection{Formal Literature}
\textcolor{black}{Three papers in formal literature discuss a handful of tools or techniques that can identify the indicators of technical debt in LLM-generated code.}

\textcolor{black}{QualAI, currently a framework under development, focuses on continuously monitoring and improving the quality of AI-based systems, with an emphasis on detecting and managing debt in ML applications throughout their lifecycles~\cite{novielli_continuous_2024}. By collecting relevant data over time, QualAI can recommend targeted changes to mitigate debt and improve overall system quality.
Velasco et al.~\cite{velasco_how_2025} present CodeSmellEval, which combines a dataset of common code smells with a new Propensity Smelly Code (PSC) evaluation specifically for LLM-generated code. Their results show that higher PSC values correlate with a greater tendency to generate code smells that cause future debt in software.
\textcolor{black}{Lastly, to mitigate deprecated API usage in LLM-generated code, another symptom of future debt in code, Wang et al.~\cite{11029774} propose ReplaceAPI, an automated approach that detects deprecated API calls in LLM outputs and replaces them with updated alternatives. This approach helps reduce future debt costs and prevents downstream technical debt caused by evolving software ecosystems.}
These frameworks offer an early method for benchmarking LLM outputs and could inform future strategies to detect and reduce different aspects of technical debt.}


\begin{table}[h]
\caption{\textcolor{black}{Tools Referenced in Grey and Formal Literature that Support Managing or Detecting Technical Debt in LLM-generated code.}}
\centering
\begin{tabular}{|p{2.5cm}|p{8cm}|p{3cm}|}
\hline
\textbf{Tools} & \textbf{Purpose} & \textbf{Open-source / Commercial} \\ \hline
SonarQube~\cite{sonarqube}, ESLint~\cite{eslint}, Pylint~\cite{pylint} & Static analysis tools to detect code smells, vulnerabilities, and performance issues & Mostly open-source (includes community editions) \\ \hline
Amazon CodeWhisperer~\cite{CodeWhisperer}, Copilot~\cite{copilot}, Tabnine~\cite{tabnine} & AI-powered coding assistants that generate code but also provide additional features such as vulnerability scanning, code quality suggestions, and in the case of Copilot, early efforts toward providing references of sources it uses & Commercial \\ \hline
Swimm~\cite{Swimm}, HopeAI~\cite{hopeai}, Graphite~\cite{graphite} & AI-powered documentation and code understanding capabilities to improve software maintainability & Commercial \\ \hline
CRken~\cite{crken}, CodeScence~\cite{codescene} & Code review automation that merges requests, flags “quick fixes” likely to add debt & Commercial \\ \hline
Codacy~\cite{codacy}, Snyk~\cite{snyk}, CAST Software~\cite{castsoftware} & AI code analysis SaaS platforms to enforce quality and security standards across the code & Commercial \\ \hline
DeepSource~\cite{deepsource}, ArmorCode~\cite{armorcode} & AI DevSecOps platform to assess code quality and SAST & Commercial \\ \hline
CodeSmellEval~\cite{velasco_how_2025} & Evaluates propensity of smelly code in LLMs & Open-source \\ \hline
QualAI~\cite{novielli_continuous_2024} & Continuously monitoring and improving the quality of AI-based systems, to manage debt in ML applications & Open-source (Under Development) \\ \hline
\end{tabular}%
\label{tab:llm-tools}
\end{table}

Table~\ref{tab:llm-tools} summarizes the tools referenced in both literatures, outlining their purposes and indicating whether they are open-source or commercial. Besides tools proposed by formal literature and static analysis tools such as ESLint, Pylint, other tools mentioned in the literature are all commercial.

\rqbox{SonarQube remains the most widely used tool for detecting technical debt indicators. Other commercial tools, such as Swimm, Snyk, and Amazon CodeWhisperer, are increasingly leveraged to detect these common risks in LLM-assisted workflows. Research contributions such as QualAI and CodeSmellEval show early efforts to create systematic frameworks and benchmarks tailored to evaluating technical debt in LLM-generated code.}

\subsection{RQ4: Are there benchmarks or datasets available to evaluate the technical debt or quality of LLM-generated code?}
\subsubsection{Grey Literature}
\textcolor{black}{Our review of grey literature revealed no standardized benchmarks or datasets currently in use for directly evaluating technical debt or its indicators in LLM-generated code. While practitioners frequently acknowledge the importance of continuous evaluation for the reliable integration of LLMs into software development cycles~\cite{B23, B17}, existing benchmarks focus almost exclusively on binary correctness. As one source emphasized, better benchmarks are needed for \textit{``preventing AI from becoming a mindless code churn machine''}~\cite{B23}. This highlights a significant gap in the availability and adoption of LLM-specific evaluation resources. There are no publicly shared or industry-standard benchmarks tailored to assessing maintainability, long-term quality, or the broader impact of incorporating LLM-generated code into software systems.}

\subsubsection{Formal Literature}
\textcolor{black}{Same as grey sources, formal literature also did not reveal any standardized benchmarks or datasets directly addressing technical debt. There is a clear gap in standardized benchmarks/datasets for technical debt in LLM-generated code. Most studies either adapt existing code smell benchmarks such as CodeXGLUE~\cite{lu2021codexglue} or CodeSmellData~\cite{velasco_how_2025} to study narrow, code-smell-related issues or rely on older datasets originally developed for assessing technical debt in software prior to the rise of LLMs~\cite{vitale_catalog_2025}.
Importantly, this limitation reflects a broader challenge in technical debt research: even for human-written code, widely accepted and comprehensive technical debt benchmarks remain scarce~\cite{avgeriou2025manifestodagstuhlperspectivesworkshop,10.1145/3345629.3345630,amanatidis_evaluating_2020}. Existing efforts are often limited in scope and tend to focus on specific debt types or proxy measures rather than long-term debt assessment.
Among formal literature, sources emphasize a growing need for better benchmarks and mention the lack of them for current LLM models~\cite{vitale_catalog_2025}.}

\rqbox{No standardized benchmarks exist in either grey or formal literature, though there is growing recognition of the need for LLM-specific benchmarks.}

\subsection{RQ5: Are there existing technical debt metrics that sufficiently capture the debt introduced by LLMs? If not, what new metrics or dimensions are being suggested?}

\subsubsection{Grey Literature}
Although no concrete new metrics are widely adopted, several sources mention the need for new evaluation dimensions and proposed early-stage alternatives for capturing the unique debt introduced by LLM-generated code.
Existing technical debt metrics, such as SonarQube rules and conventional code-quality measures, are referenced across industry discussions and are widely regarded as inadequate for capturing AI-specific debt. In multiple grey literature sources, authors noted that current metrics can surface syntax errors and some maintainability issues; they fail to capture semantic accuracy, architectural fit, and adaptability \cite{B3, B4, B54.1}.

Several sources speculate on new dimensions that could better represent LLM-induced debt. A post on LeadDev calls for context-sensitive metrics, focusing on understandability, adaptability, and reviewability~\cite{B1}, mentioning that \textit{``Although AI excels at generating one-off code, its context window is limited. Humans still play a critical role in seeing the bigger picture and understanding the full software portfolio''}. The New Stack proposes tracking post-fix behavior as a signal of long-term debt accumulation not captured by static analysis~\cite{B10}. Portkey suggests monitoring prompt reliability, token cost sprawl, and integration fragility~\cite{B22}, metrics that are novel and specific to LLM use. Other industry voices that discussed new issues, such as ethical debt, stress that these represent new concerns that call for their own dedicated metrics and evaluation approaches~\cite{B45, B53, B64}.
While SonarQube and similar tools continue to provide partial visibility, the industry recognizes that new, LLM-specific metrics are needed to capture dimensions of debt that static analysis cannot.

\subsubsection{Formal Literature}
\textcolor{black}{Only a few sources mention any new metrics for detecting or mitigating early signs of technical debt. We shortlisted five papers that either identified shortcomings in current metrics or developed their own metrics.}

Pandini et al.~\cite{pandini_exploratory_2025} combine prompt engineering and retrieval-augmented generation with Arcan, a tool for detecting architectural debt. Their evaluation shows that while such techniques improved results for smaller code debt cases, they were less effective for more complex issues, highlighting limitations of both current tooling and prompt-based mitigation.

Simoes et al.~\cite{simoes_evaluating_2024} do not strictly focus on LLM code, but do a comparison between SonarQube and LLMs for evaluating the overall quality of code. \textcolor{black}{They found that LLMs are more capable of finding finer details of quality rather than the A to E categorization used in Sonar, which shows the limitation of static tools overall. The results showed that SonarQube rates code quality based on the correcting cost of bad practices, while the LLM seems to rate the code based on the ease of reading. They also showed that there are characteristics captured exclusively by either the LLMs or SonarQube. Similarly, two other studies employ SonarQube rules to assess code quality of LLM-generated code, though they extend its use in different ways. One study complements SonarQube’s automated assessments with human evaluators, noting that judgments of quality can still be subjective~\cite{nunes_evaluating_2025}. The other also relies on SonarQube but emphasizes its limitations, motivating the need for more context-sensitive metrics that can better capture code quality across varying development scenarios~\cite{porta_prompt_2025}.}

\textcolor{black}{To find signs of code debt, such as deprecated API usages in LLMs, API Usage Plausibility (AUP) and Deprecated Usage Rate (DUR) are two metrics discussed by Wang et al.~\cite{11029774}. AUP measures how effectively LLMs predict accurate APIs without considering their deprecation status, and DUR calculates the rate of plausible completions that were annotated as deprecated. They argue that a high AUP and low DUR indicate that models predict both up-to-date and accurate APIs.}

Lastly, Foidl et al.~\cite{foidl_data_2022} focus on data debt, introducing two novel metrics: Data Smell Strength and Data Smell Density. These metrics were evaluated using rule-based and ML-based detection techniques across hundreds of datasets. This work demonstrates how new debt types, such as data quality issues, can be measured in a scalable way.

\rqbox{Grey literature mentions the need for new metrics covering dimensions such as context-sensitivity, adaptability, and reviewability. Formal literature acknowledges the limitations of static analysis tools while also exploring specific implementations such as Arcan and extensions of SonarQube, and introducing metrics like Data Smell Strength \textcolor{black}{and Deprecated Usage Rate}.}
\section{Actionable Insights and Future Research Directions}\label{sec:5}

\subsection{\underline{RQ1}: LLMs introduce both traditional and new forms of debt}
\textcolor{black}{Our results indicate a growing concern that LLMs are amplifying existing traditional technical debt and its signs in software, and introducing new types of debt, such as prompt, fast integration, and ethical debt. Both practitioners and researchers are beginning to realize that code generation with LLM systems presents a new set of challenges.} These challenges not only arise from complex code but also involve how it is produced through prompts and the quality of training data. The rapid pace and scale of integrating LLM-generated code into software systems are also raising pressing questions about how such outputs should be governed and sustained over time.

\begin{itemize}[label=\ding{43}]
    \item \underline{Future Research Directions}:
    \begin{enumerate}
        \item \textbf{Expanding taxonomies of technical debt}: Taxonomies of debt must evolve beyond traditional categories to capture LLM-specific risks. Existing debt taxonomies (e.g., Alves et al.~\cite{6974882}) are insufficient for capturing the nuanced risks of LLMs. Researchers need to formally study new categories such as prompt debt and governance debt, distinguish them from traditional code or design debt, and show how these interact. For example, our findings show that fast-integration practices can trigger cascading governance risks if left unmanaged. This gap becomes even more pressing in the context of \textit{Software Engineering 3.0}~\cite{hassan2024ainativesoftwareengineeringse}, a vision of LLM-native development characterized by conversation-oriented programming where natural language prompts serve as the primary interface between humans and AI teammates. To make this vision sustainable, the community must systematically map the unique forms of debt emerging in this new paradigm and design corresponding strategies for their detection and mitigation.
        \item \textbf{Conducting longitudinal studies}: We currently lack empirical evidence of how LLM-related technical debt accumulates and evolves over time, a gap that is particularly critical given the very nature of debt as something that emerges gradually.  
        Since AI has been increasingly integrated into software development workflows since 2021~\cite{github_study_2022, gitclear_ai_2025}, projects now routinely leverage LLMs for a wide range of tasks. This makes it an opportune time to conduct longitudinal analyses to determine whether early productivity gains are ultimately offset by growing maintenance and refactoring costs. 
        Such evidence would be invaluable for understanding the long-term sustainability and quality implications of AI-assisted software development.
        \item\textcolor{black}{\textbf{Analyzing socio-technical impacts of AI adoption in software development}: Availability of LLM-based tools is reshaping software development dynamics as these tools increasingly contribute directly to production code~\cite{li2025riseaiteammatessoftware, ehsani2026aicodingagentsfail}. We identify \textit{governance debt} as a pressing concern, highlighting the need for continuous oversight of AI-generated code to ensure quality, and despite the growing adoption of AI within software teams, studies indicate that full reliance on LLM outputs remains risky~\cite{10.1145/3696630.3730538, 11029928}. 
        Beyond code, we observe emerging challenges in \textit{knowledge management} in developer teams. Prompts and interaction patterns increasingly function as development artifacts, yet are often undocumented or scattered across tools and repositories~\cite{prompt_sprawl}. This ``prompt sprawl'' risks loss of system knowledge, weakens accountability, and complicates collaboration~\cite{prompt_sprawl}. Practices such as prompt registries, versioned documentation, and review workflows for LLM inputs are therefore essential to preserving organizational knowledge~\cite{prompt_sprawl, linkedin_code_review}. These findings highlight the need for research that moves beyond purely technical concerns to examine how AI-assisted development reshapes socio-technical aspects in software teams.}
    \end{enumerate}
    \item \underline{Actionable Insights for Practitioners}:
    \begin{enumerate}
        \item \textbf{Safeguards over speed:} Developers must not treat LLM code as ``free productivity''. Industry voices emphasize that code quality is often sacrificed for speed. Introducing safeguards such as provenance review (where did the code come from?), governance checks, and systematic prompt documentation can reduce long-term risks.
        \item \textbf{Organizational responsibility:} Beyond individual developers, organizations should institute policies that balance speed and sustainability. For instance, they could require developer review of all LLM-generated pull requests and \textcolor{black}{test} coverage before integration into the main codebase.
    \end{enumerate}
\end{itemize}

\begin{figure*}[h]
    \includegraphics[width=0.9\textwidth]{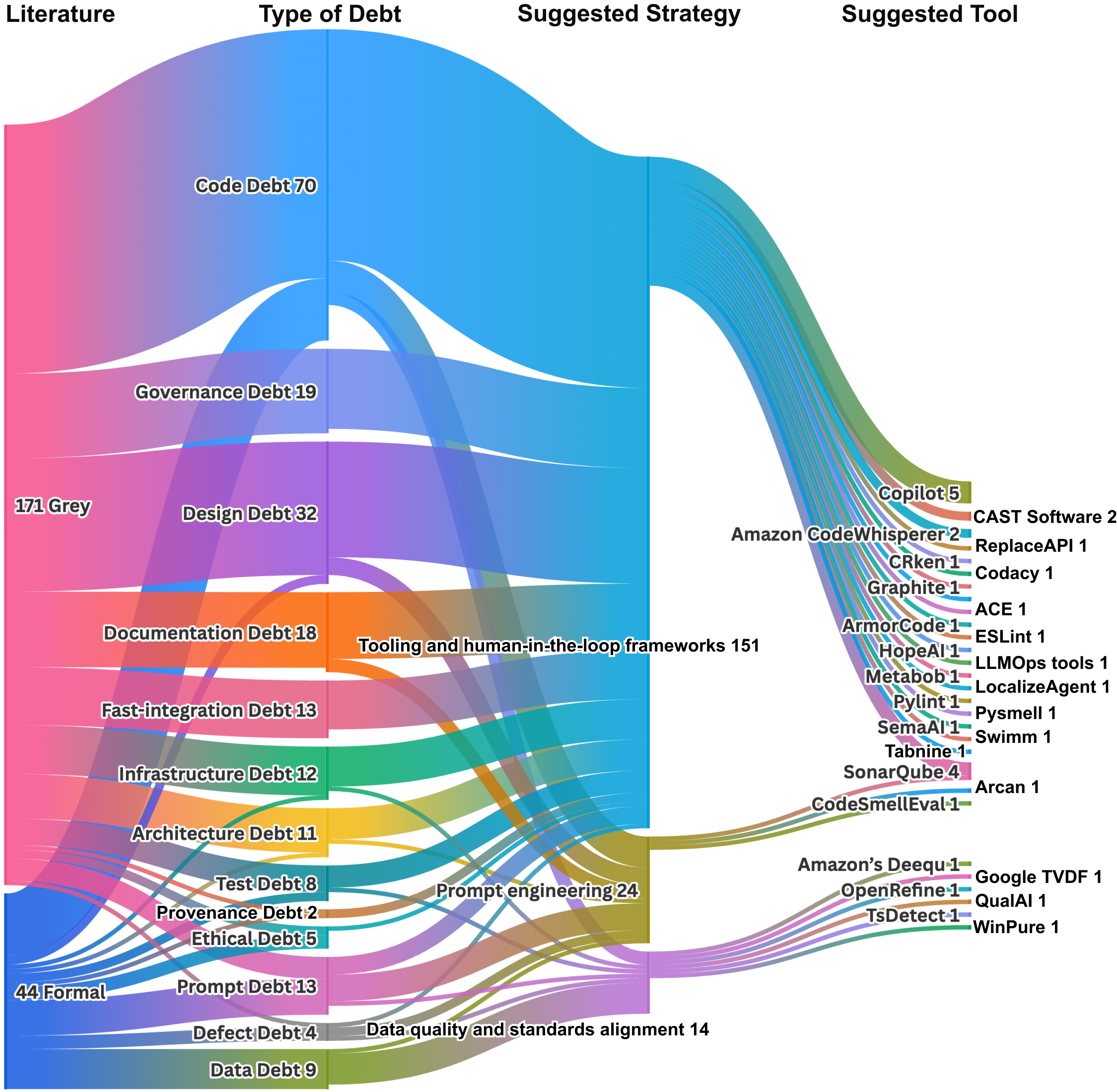}
    \caption{Mapping Technical Debt Types to Mitigation Strategies Across Grey and Formal Literature.}
    \label{fig:mapping}
\end{figure*}

\subsection{\underline{RQ2}: Strategies for addressing LLM-induced debts are unevenly distributed across debt types}
\textcolor{black}{Figure~\ref{fig:mapping} shows that code debt is the primary focus of both grey and formal literature, attracting the most concrete strategies for mitigation.} Other forms of debt, particularly new forms such as ethical, provenance, and governance debt, are acknowledged but remain disconnected from actionable practices. The strongest cluster of activity is around tooling and human-in-the-loop frameworks, which benefit from a range of commercial and open-source tools, showing the community's reliance on oversight mechanisms to tackle debt. By contrast, prompt engineering is tied to a few tools, highlighting the unresolved challenge of standardizing or automating what constitutes an effective prompt with the right amount of context. Data quality and standards alignment are similarly underrepresented, with only a handful of emerging frameworks attempting to fill the gap.

\begin{itemize}[label=\ding{43}]
    \item \underline{Future Research Directions}:
    \begin{enumerate}
        \item \textbf{Systematically characterize and measure prompt debt:} Prompting has become the central interface between developers and LLMs, and it is emerging as both a source and a solution to technical debt. Poorly designed or brittle prompts can introduce prompt debt~\cite{aljohani_promptdebt_2025}, while informal practices such as vibe coding contribute to unstable or fast-integration debt, and therefore, more governance debt in the future. In contrast, carefully crafted and well-documented prompts are often described as a key strategy for producing higher-quality, more reliable code~\cite{11025628, ehsani2025_EMSE}. Grey literature highlights this tension most directly, warning that undocumented or inconsistent prompts erode reproducibility and maintainability. Formal studies reinforce the same duality: prompting, especially when combined with techniques like retrieval-augmented generation (RAG), can mitigate design and architectural debt, yet its effectiveness diminishes in large and complex systems. This suggests that while prompting is indispensable, it remains a fragile lever for debt management that needs clearer guidelines, better tooling, and systematic evaluation.
        \item \textbf{Investigate methods for detecting data debt:} Data debt is a silent but central factor for code generation. Formal studies show that when training datasets or retrieved data in RAG systems contain noise, such as stray tokens, file paths, or non-ASCII characters, models inherit and propagate these flaws, creating downstream debt before any code is even written. This problem is compounded by the lack of transparency around commercial LLM training data, where developers often have no visibility into its quality or provenance. Recent work has begun to introduce metrics like Data Smell Density and Data Smell Strength to quantify such issues, but these remain early efforts. 
        Researchers must develop rigorous methods for detecting, quantifying, and mitigating data debt to ensure reproducibility and model trustworthiness.
        \item \textbf{Grounding ethical and provenance debts in empirical studies: }
        Ethical and provenance debts, stemming from fairness issues in LLM-assisted fixes or unclear code ownership, require stronger empirical grounding~\cite{10.1145/3774324}. Future work should investigate how opacity in data provenance and training sources introduces ethical, legal, and governance liabilities, and how these debts propagate through developer workflows.
    \end{enumerate}
    \item \underline{Actionable Insights for Practitioners}:
    \begin{enumerate}
        \item \textbf{Human-in-the-loop review systems:} 
        Teams should integrate LLM-generated suggestions into established quality assurance and debt management workflows, incorporating peer review, refactoring, and regression testing to ensure reliability and maintainability. 
        \item \textbf{Practical prompt discipline:} Practitioners should begin treating prompts as part of the development lifecycle. Versioning prompts, documenting intent, and testing prompt robustness across variations could prevent prompt debt from silently accumulating.
    \end{enumerate}
\end{itemize}

\subsection{\underline{RQ3}: Tools to detect LLM-induced technical debt are scarce and limited in scope}
\textcolor{black}{Industry predominantly relies on static analysis tools such as SonarQube to measure traditional signs of technical debt in software; however, these tools are not designed to accurately capture LLM-induced technical debt. Commercial tools such as CodeWhisperer, CRken, Swimm, and Snyk are beginning to support explainability, documentation, and maintainability in LLM-assisted workflows; however, they are typically offered under proprietary licensing models and are rarely subject to independent or rigorous external evaluation.} In contrast, academic work is only just beginning to address this gap: efforts such as QualAI, aimed at monitoring AI-generated systems, and CodeSmellEval, which benchmarks LLM outputs for code smells, mark the first steps toward building more systematic and transparent metrics for LLM-driven debt.

\begin{itemize}[label=\ding{43}]
    \item \underline{Future Research Directions}: 
    \begin{enumerate}
        \item \textbf{Developing open tooling:} Researchers must prioritize the development of transparent, open-source tools designed specifically for LLM-generated code, rather than retrofitting existing static analysis platforms. \textcolor{black}{Traditional static analysis tools are essential for ensuring code quality, but they fail to directly capture LLM-specific risks such as vague prompts, provenance ambiguity (e.g., unclear or missing source attribution), or non-deterministic outputs. New tools should aim to 
        (1) track and version prompts alongside generated code, 
        (2) measure reproducibility and reliability of LLM-generated code (e.g., assessing semantic correctness of code~\cite{sharma2025assessingcorrectnessllmbasedcode}), (3) flag potential provenance or licensing concerns of the generated output, and (4) integrate interpretability and explainability features (e.g., helping developers understand why a certain piece of code was generated ~\cite{ehsani2025bugfixingbroadercontext}). Building these tools openly would allow reproducibility and wider adoption.}
        \item \textbf{Establishing rigorous and reproducible evaluation:} Tools should be evaluated using reproducible benchmarks, enabling consistent comparisons across projects and models. Without this, tools risk remaining vendor marketing claims rather than evidence-backed solutions.
    \end{enumerate}
    \item \underline{Actionable Insights for Practitioners}:
    \begin{enumerate}
        \item \textbf{Reducing vendor dependence:} \textcolor{black}{Commercial tools provide partial coverage and require paid licenses. Teams should combine them with in-house practices for accountability, such as internal dashboards tracking LLM usage or extending existing quality pipelines with LLM-specific checks.}
        \item \textbf{Avoiding tool lock-in:} Over-reliance on vendor solutions can limit flexibility and hinder long-term sustainability. Practitioners should push for hybrid approaches (e.g., combining Snyk with open-source refactoring tools) to maintain adaptability.
    \end{enumerate}
\end{itemize}

\subsection{\underline{RQ4} \& \underline{RQ5}: Metrics and benchmarks to measure LLM-induced technical debt are missing}
\textcolor{black}{Currently, no standardized benchmarks or mature metrics exist for directly evaluating technical debt in LLM-generated code. Both grey and formal literature stress that while correctness-based benchmarks are common, they capture only short-term accuracy and overlook deeper concerns such as maintainability, adaptability, and semantic soundness. This gap prevents meaningful comparison across tools and models, leaving developers without reliable ways to assess long-term risks. Establishing LLM-specific benchmarks is therefore a critical future direction, not only to enable fair evaluation but also to guide the design of tools and practices that can genuinely mitigate LLM-induced technical debt.}

\begin{itemize}[label=\ding{43}]
    \item \underline{Future Research Directions}:
    \begin{enumerate}
        \item \textbf{Designing new metrics:} Researchers should go beyond correctness and design metrics that capture semantic accuracy, maintainability, adaptability, and even social dimensions like provenance and ethical implications. New metrics could address the \textit{social and ethical dimensions} of LLM use, such as provenance (can we trace the origin of generated snippets to licensed or unlicensed code?), reproducibility of outputs (does the same prompt yield consistent results across runs?), and fairness/ethical risk (does the generated code propagate harmful stereotypes or security vulnerabilities)~\cite{sajadi2025llms, sajadi2025safeaigeneratedpatcheslargescale}. Developing these metrics in open-source frameworks would allow integration into CI/CD pipelines, ensuring they become actionable rather than just theoretical constructs.
        \item \textbf{Developing community-driven benchmarks:} Shared, public benchmarks for LLM-generated code are essential. Similar to how CodeXGLUE~\cite{lu2021codexglue} advanced code intelligence tasks, or SWE-Bench~\cite{jimenez2024swebenchlanguagemodelsresolve} made the evaluation of LLMs easier for bug fixing, an equivalent suite would drive comparability in technical debt research.
    \end{enumerate}
    \item \underline{Actionable Insights for Practitioners}:
    \begin{enumerate}
        \item \textbf{Adopt conservative validation:} Until robust benchmarks exist, practitioners must treat LLM integration as experimental. Manual review, regression testing, and stress testing should be mandatory.
        \item \textbf{Collect organizational evidence:} Teams can build their own internal benchmarks by tracking local data (e.g., bug rates, maintenance costs of LLM code). This internal evidence base can guide decisions until standardized metrics emerge.
    \end{enumerate}
\end{itemize}

\section{Threats to Validity}\label{sec:6}

\textbf{Construct Validity: }While our source selection process involved manual filtering and is susceptible to human error, we mitigated this by having two authors independently review all sources and resolve disagreements through discussion. We also followed a rigorous quality assessment process for both formal and grey sources. For grey literature, we reviewed full content; for formal papers, we examined titles, abstracts, and full texts as needed.
\textcolor{black}{Another potential threat lies in the formulation of search strings. To address this, we tested multiple variations and selected the one that yielded the most relevant results with minimal noise.}
To reduce bias in data extraction and coding, two authors independently coded all data. For subjective coding dimensions, we calculated Cohen’s Kappa coefficients to assess inter-rater reliability, achieving high agreement scores (above 0.9). All extractions were reviewed in joint sessions with all authors to ensure consistency and accuracy.
\textcolor{black}{Study selection bias may have occurred during screening, as inclusion and exclusion decisions involve researcher judgment. To mitigate this risk, we defined explicit inclusion criteria, conducted independent screening, and resolved disagreements through discussion.}

\textbf{Internal Validity: }Our search results may have been influenced by the internal algorithms of the digital libraries and search engines used, which are not publicly documented. To mitigate this risk, we used four major academic databases (IEEE Xplore Digital Library, ACM Digital Library, Elsevier ScienceDirect, and SpringerLink Online Library) and searched grey sources through the Google search engine, increasing coverage and reducing dependence on a single indexing mechanism.
\textcolor{black}{Additionally, grey literature varies in methodological rigor compared to peer-reviewed studies. We mitigated this risk by applying a structured quality assessment checklist and excluding sources that did not meet the threshold.}

\textbf{External Validity: }\textcolor{black}{We limited our selection to English-language sources accessible through the selected platforms. This may have excluded some relevant material and affected generalizability. In addition, our reliance on the technical debt terms in the search string may bias the sample toward studies and practitioner discussions that explicitly frame quality concerns as those types of debt, while not including related work discussed under alternative terms such as maintainability, software evolution, or code quality.
To mitigate this, we used an inclusive search string of 13 established technical debts in the literature, followed by extensive manual filtering to remove irrelevant entries.} This allowed us to include as many relevant sources as possible while maintaining quality and focus.
We invested significant time in the manual inspection process, which led to the removal of over a thousand irrelevant sources.
\textcolor{black}{The included studies span diverse tools, domains, and development contexts. However, findings may not generalize uniformly across all programming ecosystems or organizational settings.}
\textcolor{black}{In addition, while our source retrieval process followed systematic guidelines and
did not favor any specific source type, a large portion of our grey literature corpus (52\%) consists of company blogs and industry reports, which reflects the current state of practitioner discourse in LLM-assisted development that may introduce a bias toward grey literature's perspectives.}
\section{Conclusion}\label{sec:7}
This study provides a comprehensive examination of the risks and challenges posed by technical debt in the context of LLM-generated code, along with strategies proposed for its detection and mitigation from both practitioner and research perspectives. Through a multivocal literature review of \textcolor{black}{104 sources}, we find that LLMs intersect with existing forms of technical debt, especially code, design, and documentation debts the most. In addition, they introduce new forms of debt, most notably data debt, fast-integration debt, and governance debt that remain underexplored and insufficiently addressed. Given the abundance of discussion around traditional debts of software from LLMs and forms of debts that are new to the special needs of LLMs, our findings emphasize the need for deeper investigation into the long-term implications of LLM use within software engineering workflows.
We highlight several strategies repeatedly discussed in both literatures, including tooling and human-in-the-loop frameworks, prompt engineering, and data quality alignment, as central approaches for managing LLM-related debt. \textcolor{black}{We also categorize several commercial and open-source tools suggested for practice, from the integration of LLMs with static analysis tools such as SonarQube and Pylint to manage early indicators of technical debt in software, to emerging LLM-assisted platforms such as Amazon CodeWhisperer, Swimm, and Copilot. However, our findings show that current solutions remain fragmented, heavily commercial in nature, and limited in their ability to fully address the breadth of LLM-related technical debt.
Finally, we identify a striking absence of standardized benchmarks and LLM-specific metrics, highlighting a major gap that limits systematic evaluation of LLM-generated code and its long-term maintainability.} Addressing this gap alongside the emerging forms of debt we highlight marks an important direction for future research, one that will be critical for ensuring sustainable and responsible integration of LLMs in software development.

In the future, research must move beyond documenting risks toward building actionable solutions. This includes designing standardized benchmarks and LLM-specific metrics, developing robust frameworks for prompt management and evaluation, and creating open, community-driven datasets that reduce reliance on commercial black-box models. Equally important is studying the temporal evolution of LLM-related debt, to understand not only how such debt emerges but also how it accumulates or can be repaid over time. By pursuing these directions, the software engineering community can better align research, practice, and tooling to manage technical debt in the era of LLMs.

\bibliographystyle{ACM-Reference-Format}
\bibliography{ref,MLR}

\end{document}